\documentclass{article}
\usepackage[english]{babel}
\usepackage[applemac]{inputenc}
\usepackage{geometry}
\usepackage{graphicx} % Add graphics capabilities
\usepackage{booktabs} % ``Proper'' table layout
\usepackage{amsmath}  % Better maths support
\usepackage{amssymb} 
\usepackage{amsthm}
\usepackage[style=numeric, backend=bibtex]{biblatex}

\usepackage[colorlinks=true,linkcolor=red]{hyperref} % Hyperlink capabilities
\usepackage{array}
\newcolumntype{C}[1]{>{\centering\arraybackslash}p{#1}@{}}
\usepackage[font=scriptsize]{caption}
\usepackage{tikz}
\usetikzlibrary{arrows}
\usetikzlibrary{automata}
\usepackage{bbm}
\usepackage{bm}
\usepackage{abstract}
\usepackage{mdframed}
\usepackage{lineno}
\usepackage{comment}

\begin{document}

\title{Employing Gaussian process priors for studying spatial variation in the parameters of a cardiac action potential model}

\author{ Alejandro Nieto Ramos$^{1}$, Christopher L. Krapu$^{2}$,\\Elizabeth M. Cherry$^{3}$, Flavio H. Fenton$^{4}$\\
$^1$ School of Mathematical Sciences,\\Rochester Institute of Technology,\\
$^2$ Geospatial Science \& Human Security Division, Oak Ridge National Laboratory,\\
$^3$  School of Computational Science and Engineering,\\Georgia Institute of Technology,\\
$^4$  School of Physics,\\Georgia Institute of Technology. }

\maketitle

\begin{abstract}
Cardiac cells exhibit variability in the shape and duration of their action potentials in space within a single individual. To create a mathematical model of cardiac action potentials (AP) which captures this spatial variability and also allows for rigorous uncertainty quantification regarding within-tissue spatial correlation structure, we developed a novel hierarchical Bayesian model making use of a latent Gaussian process prior on the parameters of a simplified cardiac AP model which is used to map forcing behavior to observed voltage signals. This model allows for prediction of cardiac electrophysiological dynamics at new points in space and also allows for reconstruction of surface electrical dynamics with a relatively small number of spatial observation points. Furthermore, we make use of Markov chain Monte Carlo methods via the \texttt{Stan} modeling framework for parameter estimation. We employ a synthetic data case study oriented around the reconstruction of a sparsely-observed spatial parameter surface to highlight how this approach can be used for spatial or spatiotemporal analyses of cardiac electrophysiology.
\end{abstract}

\section{Introduction} 

Mathematical models can be used to improve our understanding of real-world phenomena. As a simplification of a physical or engineered system, a model can provide insights into observed behavior, describe patterns, or make predictions. While the form of the model equations can express qualitative relationships among the system variables, well-chosen parameter values are needed for a quantitative understanding. However, appropriate selection of parameter values can be challenging for systems that exhibit variability. Additional assumptions regarding the heterogeneity of model parameters in space or time may help provide a more faithful and less restrictive representation of the system of interest.
In biological system, uniformity is relatively rare and variability occurs naturally across all levels, from the microscopic to the macroscopic. Cardiac cells in particular exhibit strong variation in the shape and duration of the electrical signals that trigger muscle contraction, not only across individuals, but also in space and time within a single individual.

%I struck out the following line because we don't want to confuse the reader as to whether or not we focus on cross-individual variation.
%The presence of variability makes it difficult to predict outcomes, such as the effects of a drug or treatment on a population, because different individuals may exhibit different behavior. 

Some aspects of natural variation can be represented in a model through the use of nonstationary parameter sets that describe separate individuals. However, preexisting models of cardiac action potentials avoid parameter estimation for a spatially varying set of parameters and are often employed using a single value for the entire spatial domain. It would be highly advantageous to be able to describe variability more systematically; towards this end, we would like to characterize admissible parameter sets that can represent variability across individuals or even within the same individual over time or space. In this paper we aim to represent variability of cardiac action potentials in space for one individual.

\subsection{Cardiac electrophysiology}

%@Alejandro: this isn't suitable for a research paper - if you are going to make a citation, use peer reviewed work and don't just say "according to the CDC".

%Heart disease is the leading cause of death in the United States, causing about 1 in 4 deaths \cite{Murphy2018}. The study of cardiac electrophysiology is pertinent because cardiac electrophysiology models are able to reproduce alternans, which are known to be precursors of arrhytmias. In turn, they can lead to sudden cardiac arrest or stroke. \al{I feel this part could be out of place}

\subsubsection{Cardiac action potentials}\label{CAP}

The electrical dynamics of cardiac tissue are governed by the aggregate behavior of large numbers of excitable cardiac myocytes. These cells may be stimulated electrically either by the sinoatrial node in nature or via external probes in an experimental setup.If the applied stimulus increases the cell's membrane potential above a threshold value there is a difference in voltage, which is captured by measurements of the cell's action potential (AP). When this occurs , alterations to the cell's membrane potential induces ions to transit through membrane, aided by specialized proteins called porins and thereby inducing an electric current. 

\begin{figure}[h]
	\centering
	\includegraphics[angle=0, width=1\textwidth]{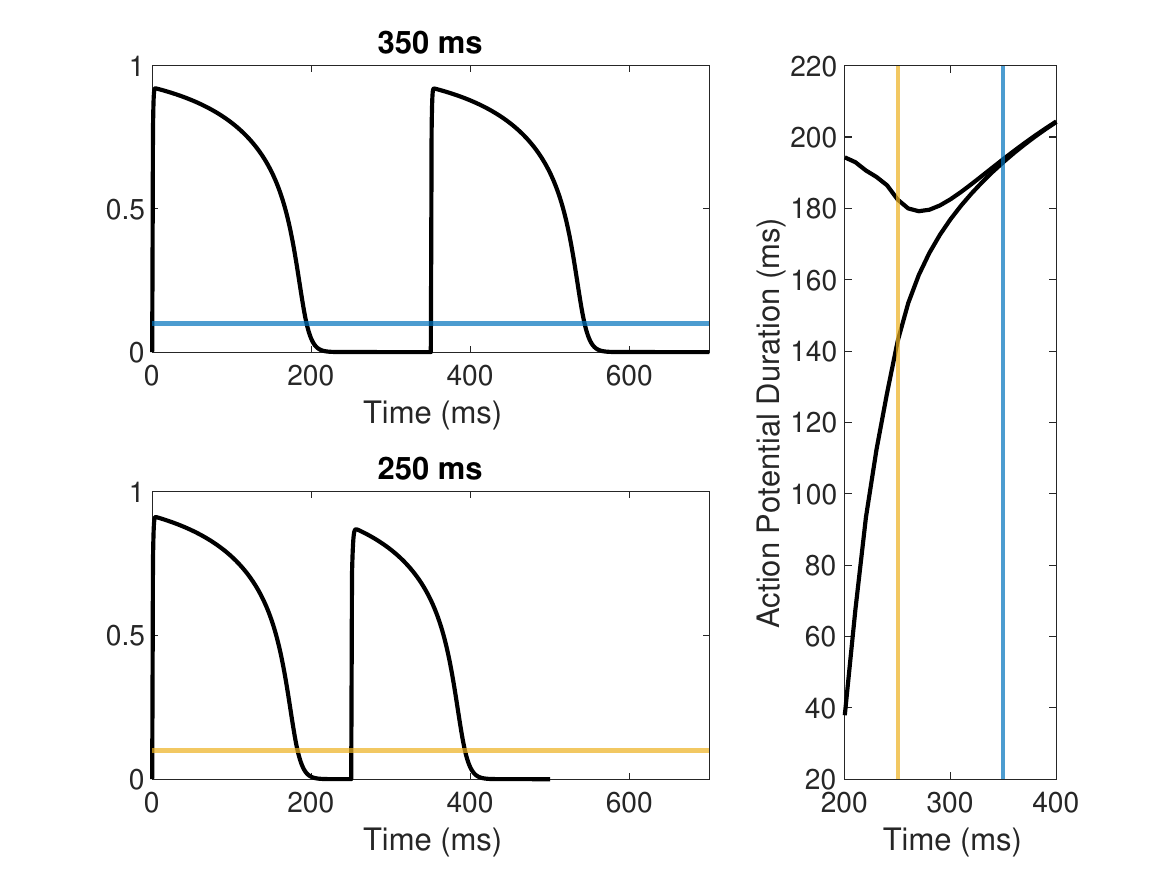}
	\caption{\label{alternans} Action potential duration as a function of cycle length using the Mitchell-Schaeffer model introduced below. The left panel shows action potentials at CLs of 350 and 250 ms., with alternans present for the shorter CL. Horizontal lines correspond to APDs at 90\%. %\chris{We'll need to rework this figure. I recommend moving the inner plots to the side and stacking them on top of each other, so we have one large plot on the left and two stacked smaller plots on the right. Remember to put the time units on the y-axis label as well.}
	}
\end{figure}

The time that it takes for an action potential to depolarize and repolarize given a threshold-exceeding stimulus is known as the \emph{action potential duration} (APD). Under natural conditions, these action potentials exhibit oscillatory behavior, eventually converging to steady-state electrical dynamics. After reaching steady state, the action potential duration between APs is constant. However, if the pacing becomes faster, the APD alternates between long and short values, a phenomenon called \emph{alternans}. These dynamics can be represented by a \emph{bifurcation plot} (BP), where APD is graphed against period, also known as cycle length (CL). In Figure \ref{alternans} we see two APs paced at 350 and 250 milliseconds; the the shorter cycle length also exhibits alternans in this case. This behavior is also represented in a BP where two branches are present when alternans appear.

\begin{figure}[h]
	\centering
	\includegraphics[angle=0, width=1\textwidth]{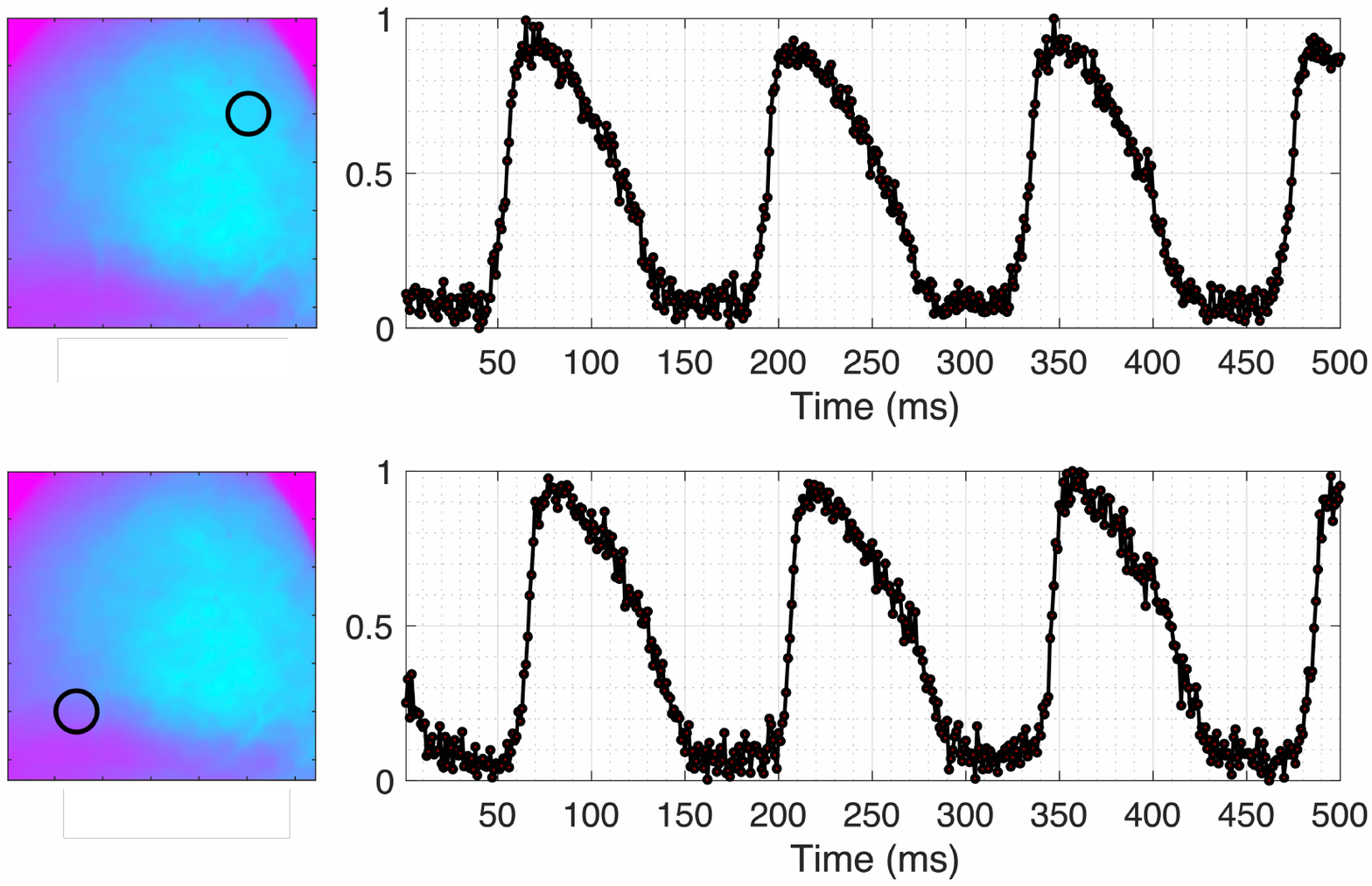}
	\caption{\label{OM} Voltage mapping obtained with optical mapping from a rabbit heart paced at 280 ms. (left), where two series of action potentials are shown at two different locations represented by the center of the black circles.
	%\al{I think we really need this graph, because people in general do not know what optical mapping is. I even used the same colors that we are using for the grids. I have been thinking of where to present this part, but I am not sure yet.}
	}
\end{figure}

Characterizing the spatiotemporal dynamics of myocyte action potentials is of great scientific interest; the measurement technique of optical mapping is frequently used towards this end. Optical mapping is a method used to directly record the electrical activity of the heart at high spatio-temporal resolution through the use of fluorescent dyes. It allows for mapping the propagation of electrical signals across the myocardium, thereby allowing characterization of electrical dynamics in the two-dimensional plane (Figure \ref{OM}). After these signals are recorded, action potential durations may be calculated and mapped in order to visualize the variability of action potentials in space at different CLs.

\subsubsection{Modeling cardiac action potentials}
Denis Noble developed the first cardiac action potential model in 1962 to simulate the action potential dynamics of Purkinje cells \cite{noble1962}. This model was itself a modification of a previous mechanistic model created by Hodgkin and Huxley \cite{hodgkin1952} for the neuronal action potential of a squid axon. Since then, several types of models have been developed, representing different kinds of cells from diverse species. %\chris{Some more citations would be nice here} 
Over time, the models have become more complex, usually including more ionic currents, and as a consequence, more differential equations and parameters. They have also been modified to include cell communication and and spatiotemporal calcium cycling dynamics. Depending on what is being studied, phenomenological models can be used versus more detailed models developed from first principles \cite{Qu2014}.

\subsubsection{Parameter estimation and prediction}

Several approaches have been used previously to describe variability in the context of cardiac electrophysiology and related fields. Britton et al.~\cite{Britton2013} pioneered the use of a population of models constructed to represent variability at the population level, with the goal of providing a tool for understanding and quantifying variability in the response to drugs. To do so, they selected a small subset of model parameters and generated random combinations of parameter values sampled from the large parameter space. Candidate parameterizations were integrated into the model to generate quantities of interest, such as action potential duration, for comparison with reference values. Only those parameter sets that satisfied all specified criteria were included in the final population.
S\'anchez et al.~\cite{Sanchez2014} followed the same methodology but extended its use to three different models in order to study how strongly the choice of model influenced conclusions that could be drawn regarding the role of specific model parameters and components on behavior. 

Sarkar et al.~\cite{Sarkar2012} followed a different approach by treating sets of parameter values as inputs and quantities of interest (like APD) resulting from those sets as outputs, then using multivariable linear regression~\cite{Sarkar2010} to relate the two. The process yielded a matrix of parameter sensitivities that indicated how much a change to each parameter affected each quantity. If the sets of output quantities are constrained to realistic ranges of values, this approach can provide detailed quantitative information about variability and sensitivity within the population.

In yet another approach, Pathmanathan et al.~\cite{Pathmanathan2015} used a nonlinear mixed effects model to capture average behavior along with a term to represent individual variation in cardiac sodium channel inactivation. Each data set then could be fitted individually while also maintaining an average-cell fit that was not simply the average of the fitted data; using such an average produced characteristics that were not consistent with their data. Uncertainty quantification tools then were applied to understand the effects of variability on outputs including APDs. 

These previous approaches to representing variability in cardiac cell electrophysiology have some limitations. The population approach is computationally inefficient because the vast majority of generated candidate parameterizations are not used. Most of the studies use a limited amount of experimental data, such as summary-level biomarkers (e.g., APD) or components of current descriptions such as sodium channel inactivation rather than a time series of system variables such as the voltage, which may make lower-level predictions about action potential shape and duration as well as the relative roles of particular transmembrane currents less reliable. In addition, all of the approaches suffer from performance limitations as more parameters are added. Finally, although these approaches all try to describe differences across a population, none of them generates the actual probability distributions of parameter values given specific input data; at best, they generate approximations of these distributions. 

Related work has been done to apply automatic optimization to reconcile true and modeled cytosolic calcium transient traces and APs for the Severi-Di Francesco CAP model \cite{Fabri2017} as well as intracellular calcium and sodium concentrations \cite{Krogh2017}. Least-squares optimization methods have been used to characterize nonidentifiability for the Luo Rudy phase I equation system \cite{Zaniboni2010} and to perform parameter estimation using the Beeler-Reuter ionic equations for the ventricular action potential \cite{Dokos2004}. Other relevant studies have looked at parameter sensitivity analysis in electrophysiological models via  multivariate regression \cite{Sobie2009}, parameter estimation under multiple AP models \cite{mann2016}, and parameter estimation via genetic optimization \cite{Cairns2017}.

\begin{comment}
\begin{itemize}
\item  Applied automatic optimization to reduce model discrepancy between experimental data and the Severi-Di Francesco cardiac action potential model and to fit recorded APs and cytosolic calcium transient traces.

\item \cite{Dokos2004} Used Least-squares optimization by curvilinear gradient to perform parameter estimation using the Beeler-Reuter ionic equations for the ventricular action potential.

\item \cite{Sobie2009} used multivariate regression to study parameter sensitivity analysis in electrophysiological models.

\item \cite{Zaniboni2010} Applied a least square iterative method on the Luo Rudy phase I equation system to derive two different sets of parameters that generate two almost identical AP profiles. 
 
\item \cite{mann2016} used a global optimization approach to improve three existing different action potential models and reproduce clinical data sets more closely by recalibrating the models.  

\item \cite{Krogh2017} Using a human ventricular myocyte model, they show that model optimization to clinical long QT data constrains model parameters in a satisfactory manner using certain bounding intracellular calcium and sodium concentrations.

 \item \cite{Cairns2017} used a genetic algorithm to parameterize cardiac action potential models.
\end{itemize}
\end{comment}

\subsection{Bayesian methods for parameter estimation}

%\subsubsection{Bayesian nonlinear ODE models}

%In \cite{ramsay2007}, a review of methods to parametrize ODEs are presented, and some of them are Bayesian.

\subsubsection{Bayesian algorithms for parameter estimation}
While the optimization-centric approaches in the previous section yield several advantages including simplicity and relatively quick execution, they are less well-suited for uncertainty quantification and probabilistic modeling. For these purposes, a statistical modeling framework may be desirable. We adopt a Bayesian approach in this study as it allows for two primary reason; first, by using Monte Carlo methods, we are able to freely use any model of choice for our analyses without functional or distributional restrictions. Second, this approach allows us to conduct principled uncertainty quantification (UQ) via characterization of the posterior distribution over the model parameters given the data and appropriately specified prior distributions over model parameters. We perform this characterization using Markov chain Monte Carlo (MCMC) to draw samples from the posterior distribution. The objective of MCMC methods within the context of Bayesian modeling is to design a Markov chain in such a way that the stationary distribution of the chain coincides with that of the posterior distribution. In the case of parameter estimation for an ordinary differential equation (ODE), the interest lies in finding the distribution of parameters used as inputs for the ODE model. In \cite{ramsay2007}, a review of methods to parametrize ODEs are presented, some of which are Bayesian.

A diverse range of MCMC methods have been developed, some of which are widely-applicable with relatively little problem-specific information, while others require bespoke implementations for each probability model considered. As used in statistical inference, they are employed to provide sample-based summaries of a posterior distribution over model parameters.
The Metropolis-Hastings algorithm \cite{metropolis1953, hastings1970} is perhaps the simplest method and requires only evaluating the model's unnormalized posterior density a single time for each MCMC proposal. A common choice with this method when used for real-valued inference problems is to use a single proposal distribution such as a multivariate Gaussian to generate new proposal candidates. Gibbs sampling \cite{gelfand1990} is another popular MCMC method and is applied by iteratively selecting a single parameter and making a draw conditional of values of all other parameters. While these methods are quite general, they do not make use of gradient information from the posterior and may exhibit random walk behavior which slows convergence to the posterior distribution \cite{neal2011}. As a consequence, it may be advantageous to use gradients of the log-posterior for models with a continuous parameter space. Hamiltonian Monte Carlo \cite{duane1987, neal2011} makes Monte Carlo proposals guided by these gradient terms and exhibits much more rapid convergence to the target distribution \cite{beskos2013, mattingly2012}. An alternative paradigm is to identify a simplified surrogate distribution for the posterior distribution and employ optimization to minimize a variational loss function. Generic methods to do this for arbitrary probability models with continuous parameter spaces are now widely available \cite{kucukelbir2016} as a complement to MCMC methods. If the likelihood is unknown or intractable, approximate methods like approximate Bayesian computation \cite{Marin2012} can be used for parameter estimation.

\subsubsection{Applications of Bayesian modeling in electrophysiology}

The usage of Bayesian modeling in electrophysiological studies has a rich history extending several decades into the past and is catalogued in a review paper and book chapter \cite{barr2010, macke2015}. A common theme in these reviews is the extensive usage of prior information to help stabilize or regularize model inversion techniques. These can, for example, expedite estimation of signal source locations within cardiac tissue provided noisy observational data of surface potentials.

Bayesian approaches have been used previously with cardiac models for parameterization \cite{Coveney2018,Daly2015,siekmann2011} and parameter sensitivity \cite{Chang2015, Coveney2018}. In particular, Gaussian process priors have been used to parametrize \emph{emulators} which are sometimes used as a drop-in replacement for ODE dynamics \cite{ghosh2018}. We use them differently in this work to provide a prior for parameters of the model, rather than for the model function itself. Additionally, Bayesian techniques have not been used to study spatiotemporal variability. We will describe our approach based on obtaining distributions of model parameters using Bayesian methods to better characterize variability in cardiac electrophysiology and provide action potential predictions making use of spatial correlation patterns.

\section{Methods}

\subsection{Cardiac Action Potential Model}

We use the Mitchell-Schaeffer (MS) model \cite{mitchell2003} to fit noisy synthetic [and experimental] data. MS is a phenomenological model that describes the dynamics of the cardiac membrane. It contains two variables, the membrane voltage and a gating variable and it has 5 parameters. The equations for this model can be obtained from equations of the Karma model introduced in 1994.

\begin{align*}
\frac{du(t)}{dt}&= I_{in}+I_{out}+I_{stim},\\
\frac{dh(t)}{dt}&= \left \{ 
			\begin{array}{rl}
  				\frac{1-h}{\tau_{open}}, & u < u_{gate}\\
				\frac{-h}{\tau_{close}}, & u > u_{gate},\\
 			\end{array} \right.	
\end{align*}
where
\begin{align*}
I_{in}&= h \frac{u^2(1-u)}{ \tau_{in} }\\
I_{out}&= - \frac{u}{ \tau_{out} }. \\
\end{align*}

The variable $u$ represents the voltage or transmembrane potential and $h$ is a gating variable. $I_{in}$ is the inward current and combines all the currents which rise the voltage across the cell membrane; $I_{out}$ combines the currents which decrease the membrane voltage. The stimulus current $I_{stim}$ is an external current that the experimenter applies in short pulses. %For notational consistency with existing work, we later refer to the external stimulus as the variable $\bm{x}$ though it is identical to $I_{stim}$ in this application. 
%\al{Chris, I am confused about the stimulus named $\bm{x}$, since the ODE system is defined as $ \bm{x}'=f(\bm{x},t;parameters)$}
%\chris{Then go ahead and change the notation to something sensible - just make sure it's consistent throughout the document"}

\subsection{Computational implementation}

The Heaviside function on the gate variable function $h$ can be replaced by a smooth function. This change helps when making inference with Hamiltonian Monte Carlo \cite{neal2011} in a platform like Stan \cite{stan2022}, since the model's log prior density and log-likelihood must be differentiable in all model parameters.

\begin{align*}
\frac{dh(t)}{dt}&= \frac{1-h}{\tau_{open}} \left[ 1-p \right]+ \frac{-h}{\tau_{close}} p\\
\text{where} \quad
p&=\frac{1}{2}\left( \tanh(k(u-u_{gate}))+1\right)\\ 
\end{align*}

The parameter $k$ which controls the steepness of the hyperbolic function---the jump's speed---can be fitted easily and $k=50$ gives a good approximation for the Heaviside function.

%\al{Now this paragraph has been replaced for a longer introduction}
%Another important aspect of electrophysiological cardiac models like Mitchell-Schaeffer is that when fitting synthetic or experimental data, two consecutive action potential profiles need to be fitted, since alternans could be present. Alternans is a phenomenon that occurs when applying a stimulus each specific period or cycle length. When an excitable cardiac cell---a myocyte---receives a sufficiently strong stimulus, the membrane potential changes, which in turn causes a difference in voltage represented by an action potential (AP). After reaching steady state, the time that it takes for an action potential to depolarize and repolarize given a threshold---the action potential duration---between APs is constant, but if the pacing becomes faster, the APD alternates between long and short. This can be represented by a bifurcation plot, where APD is graphed against cycle length (see Fig. \ref{alternans}).  This implies that cardiac models need to be fitted using more than one cycle length, using cycle lengths close to the bifurcation point, for which alternans are both present and absent. 
%@alejandro: you are reusing the same symbol (for example, you use x as both an ODE state variable and also a spatial coordinate for the GP). Make sure all of these symbols are unique.

\subsection{Hierarchical probability model}

Given the system defined by MS 
\[ \bm{x}'=f(\bm{x},t;\tau_{in},\tau_{out},\tau_{open},\tau_{close},u_{gate}),\] where $\bm{x}(t) \in R^2$ and $t \in \mathbb{R}^+$, we consider
\[y_{ij}=u(t_{ij};\tau_{in},\tau_{out},\tau_{open},\tau_{close}^{(i)},u_{gate})+\epsilon_{ij}\]
where $i=1,2,\ldots,N$ names the series of action potential profiles to be fitted at position $\bm{s}_i$, $u=x_1$ is the solution to MS representing the voltage, $j=1,\ldots,T$ is the number of time points for each position and $\epsilon_{ij}\sim \mathcal{N}(0,\sigma_{i})$.

We define $\bm{\lambda}^{(i)}=(\tau_{out}^{(i)},\tau_{open}^{(i)},\tau_{close},u_{gate}^{(i)})$ and $\tau_{i}$ as the parameter $\tau_{in}$ at each position $\bm{s}_i$ in a spatial domain $\mathcal{D}\subset \mathbb{R}^2$. We assume that given the position, the measurement error is the same at all times ($\epsilon_{i}\sim \mathcal{N}(0,\sigma_i)$). Let $\bm{U}$ be a matrix of size $N \times T$ where row $i$ is of the form
\[(u(t_{i1};\tau_i,\bm{\lambda}^{(i)},\sigma_{i}),\ldots,u(t_{iT};\tau_i,\bm{\lambda}^{(i)},\sigma_{i}))\]%\bm{u}_i(\bm{r};\tau_i,\bm{\theta}) \text{ where } \bm{r}=(t_1,t_2,\ldots,t_T).\]

%We assume that given the position, the measurement error is the same at all times $\epsilon_{i}\sim \mathcal{N}(0,\sigma_i)$; therefore,  $\bm{\theta}_i=(\tau_{out},\tau_{open},\tau_{close},u_{gate},\sigma)$. 

\begin{figure}[]
	\centering
	\includegraphics[angle=0, width=0.5\textwidth]{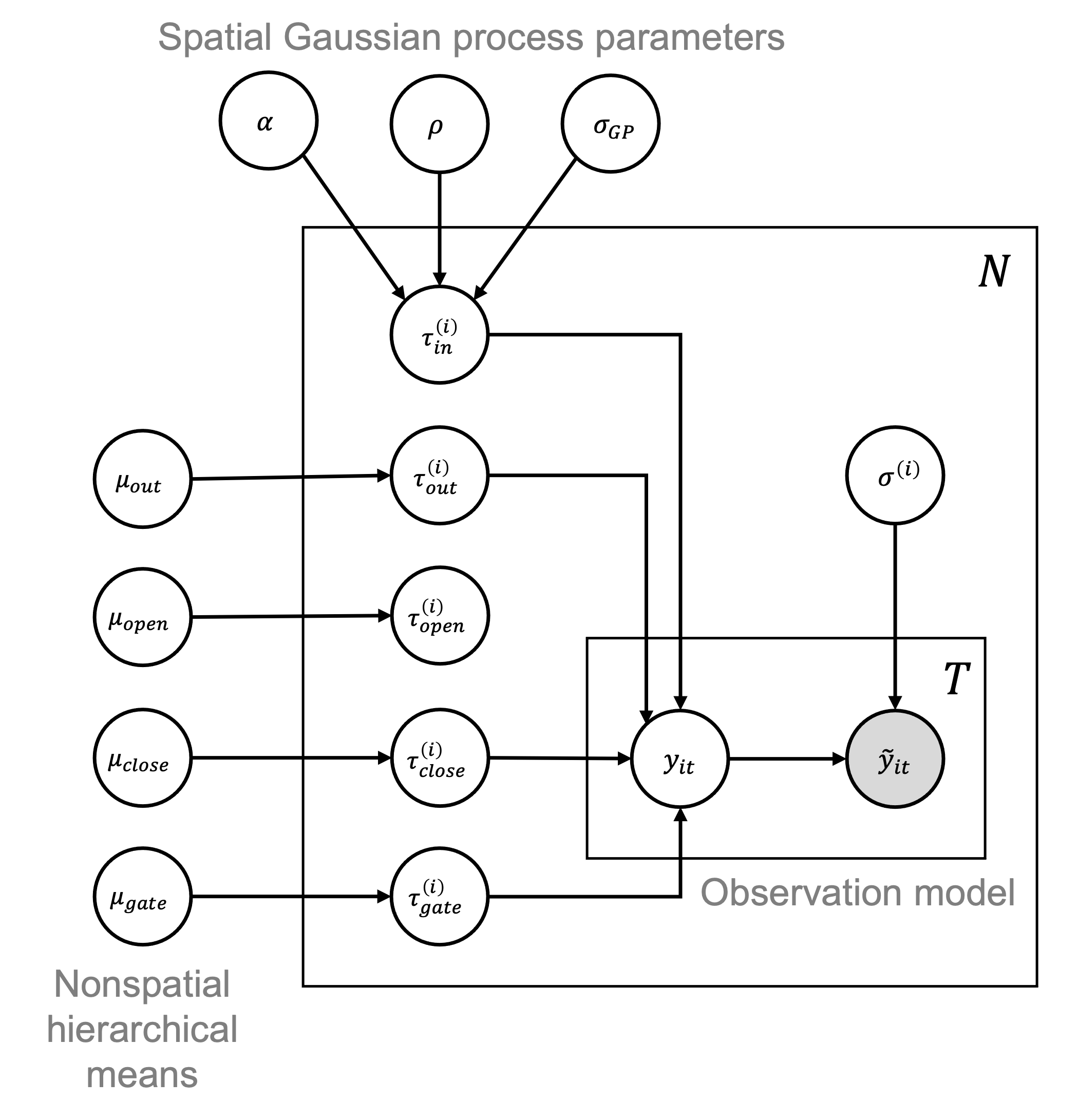}
	\caption{\label{hierarchical} Probability model in plate notation. Circles indicate random variables while boxes (plates) indicate repetitions. We consider $N$ spatial points, each of which has $T$ observations associated with it. Hollow circles are unobserved or latent, while filled-in circles correspond to observational data. }
\end{figure}

We formulate the inference problem as a hierarchical Bayesian model for the spatiotemporal voltage measurement $y_{ij}$. Our strategy is to instantiate one separate instance of the MS model at each location and use it to relate spatial variations in parameter values to resulting outputs in spatiotemporal voltage signals.
%\al{the action potential at position $i$ is $y_i$; Can we call it spatiotemporal though?}
%\chris{Yes - I've added some more clarifying text here to indicate how the problem is spatiotemporal. The voltage is indexed by both time and space.}
For the parameter $\tau_{in}$ of the MS model, we use a Gaussian process prior to encode spatial autocorrelation amongst the values of $\bm{\tau}=(\tau_{1},...,\tau_{N_1})$ used to parameterize the individual AP models at locations $\bm{s}_1,...,\bm{s}_N$. The specification for the model's parameters is summarized as

\begin{align*}
	%y_{ij}&=u(t_{ij};\tau_{in},\tau_{out},\tau_{open},\tau_{close},u_{gate})+\epsilon_{ij}\\
	\epsilon_{i}&\sim \mathcal{N}^+(0,0.5)\\
    \mu_k &\sim \mathcal{N}^+(\text{tv}(\lambda_k^{(i)}), 0.2*\text{tv}(\lambda_k^{(i)})),  \, k=1,2,3,4 \\
    %\mu_2 &\sim \mathcal{N}^+(\text{tv}(\tau_{open}), 0.2*\text{tv}(\tau_{open})) \\
    %\mu_3 &\sim \mathcal{N}^+(\text{tv}(\tau_{close}), 0.2*\text{tv}(\tau_{close})) \\
    %\mu_4 &\sim \mathcal{N}^+(\text{tv}(u_{gate}), 0.2*\text{tv}(u_{gate}))\\
      \lambda_k^{(i)} &\sim \mathcal{N}^+(\mu_k, 0.2*\text{tv}(\lambda_k^{(i)})), \, k=1,2,3,4 \\
     %\theta_2 &\sim \mathcal{N}^+(\mu_2, 0.2*\text{tv}(\tau_{open})) \\
     %\theta_3 &\sim \mathcal{N}^+(\mu_3, 0.2*\text{tv}(\tau_{close})) \\
     %\theta_4 &\sim \mathcal{N}^+(\mu_4, 0.2*\text{tv}(u_{gate})) \\
      \bm{\tau} &\sim \bm{\mathcal{N}}(\bm{\mu}^{(1)}, \bm{K}_{\alpha, \rho}+\sigma_{GP}\bm{I}_{N_1})\\
   \alpha &\sim \mathcal{N}^+(0, 1)\\ 
    \rho &\sim \text{Uniform}(1,b),\\ 
    \sigma_{GP} &\sim \mathcal{N}^+(0, 1).
\end{align*}

where $\mathcal{N}^+$ represents a folded normal distribution, tv$(\cdot)$ the value used to generate the synthetic data (\emph{true value}), $\mathbb{I}_{N_1}$ the $N_1 \times N_1$ identity matrix $b$ is the maximum distance between any two positions on the grid (those in opposite corners of the grid). An schematic of the hierarchical model can be seen in Fig. \ref{hierarchical}.

%We implement a spatially informed prior distribution for $\bm{\tau}$ at locations $\bm{s}_1,...,\bm{s}_N$ by considering these parameter values to be realizations of a random function $\tau(\bm{s})$ which has
More specifically, the Gaussian process prior is

\begin{equation}
\tau(\bm{s}) \sim \mathcal{GP}(m(\bm{s}), C(\bm{s}, \bm{s}'; \rho, \alpha))
\end{equation}

where $m(\bm{s}): \mathbb{R}^2\rightarrow \mathbb{R}$ denotes a mean function and $C:\mathbb{R}^2\times \mathbb{R}^2\rightarrow \mathbb{R}$ denotes a suitable covariance kernel function parameterized by a spatial correlation distance parameter $\rho$ and a scale parameter $\alpha$. We specify the mean function as a constant with a value of $m(\bm{s}_i)=0$ for all $i=1,2,\dots,N_1$. Then, the covariance function gives rise to the covariance matrix $K$ via the construction

\begin{equation}
\bm{K}_{\alpha, \rho} = \begin{bmatrix} 
    C(\bm{s}_1, 
\bm{s}_1) &  C(\bm{s}_1, 
\bm{s}_2) & \dots \\
    \vdots & \ddots & \\
     C(\bm{s}_1, 
\bm{s}_{N_1}) &        &  C(\bm{s}_{N_1}, 
\bm{s}_{N_1})
    \end{bmatrix}
\end{equation}

In this work, we use the squared exponential covariance function, defined as 
\begin{equation}
C(\bm{s}, 
\bm{s}'; \rho, \alpha) = \alpha^2 \exp \left( \frac{ || \bm{s} -\bm{s}'||^2}{2\rho^2} \right)
\end{equation}

where $|| \bm{s} - \bm{s}'||$ denotes the Euclidean distance between two locations $\bm{s}$ and $\bm{s'}$. The role of the correlation distance parameter $\rho$ is to decide roughly how far spatial correlations should extend, while the parameter $\alpha$ determines the relative scale of that variation. %We note that this is a standard presentation of Gaussian process priors; we refer the reader to \al{the reference is missing} for more detail. 
Alternative choices of covariance kernels include the Matern or rational quadratic kernels. If additional spatial structure such as oscillations or trends in space are present, sinusoidal and linear kernels could also be employed. It is important to note that calculating the prior density associated with $\log \bm{\tau}$ requires inversion of the pairwise covariance matrix $\bm{K}_{\alpha, \rho}$ which can be numerically unstable. To circumvent this issue, we make use of the Cholesky decomposition $\bm{K}_{\alpha, \rho} = \bm{LL}^T$ within \texttt{Stan}.\\
\par

%\al{Chris, starting here the notation does not agree with what we defined before. $\bm{\theta}=(\tau_{out},\tau_{open},\tau_{close},u_{gate})$ and $\bm{\tau}=(\tau_1,\ldots,\tau_{N_1})$ had been defined before. I also do not know what $m,\bm{X}$, and $\lambda$ are. Finally, I think instead of $f_{\bm{\tau}_i}(\bm{x}_i, t)$ it should be $y_{ij}$. There are two kinds of sigmas: $\sigma_i, i=1,ldots,N_1$ and $sigma_{GP}$.}
%\chris{The "m" is so that we can compactly write down the nonspatial parameters by indexing over them. With regard to the notation, just make sure that the parameters that you have in your Stan model are present in this equation in some way. Remove characters as you see fit.}

In total, the vector of free parameters $\bm{\theta}$ for the probability mode for all $N_1$ spatial locations is 

\begin{equation*}
     \bm{\theta} = (\bm{\tau},\bm{\lambda},\bm{\sigma},\alpha, \rho, \sigma_{GP})
\end{equation*}
where we let $\bm{\sigma}=(\sigma_1,...,\sigma^2_{N_1})$ denote a vector of measurement noise variance parameters concatenated together and $\bm{\lambda}=(\bm{\lambda}^{(1)},...,\bm{\lambda}^{(N_1)})$. Then, with the abbreviation $\left[\cdot\right] = p(\cdot)$, the prior distribution as described previously factorizes as
\begin{equation*}
     p(\bm{\theta}) = \underbrace{\left[\bm{\tau} \middle| \alpha, \rho, \sigma_{GP} \right] \left[\alpha\right] \left[\rho\right]\left[\bm{\sigma}\right]}_\text{Spatial parameters} \underbrace{\prod_{k=1}^4 \prod_{i=1}^{N_1} \left[\bm{\lambda}^{(i)} \middle| \mu_k \right] \prod_{k=1}^4 \left[ \mu_k\right]}_\text{Nonspatial hierarchical prior}.
\end{equation*}
where we let the index $k\in {1,...,4}$ run over the non-spatially varying parameters. Finally, by assuming independent Gaussian measurement error variates which are identically distributed within spatial locations, we can write the likelihood and its log-transformation as 
\begin{align}
    p(\bm{Y} \vert \bm{\theta}) &= \prod_{i=1}^{N_1} \prod_{j=1}^T \frac{1}{\sqrt{2\pi\sigma^2_i}} \exp \left[ \frac{-(y_{ij} - U_{ij})^2}{2\sigma^2_i} \right]\\
    \log p(\bm{Y} \vert \bm{\theta}) &= \frac{1}{2} \sum_{i=1}^N \frac{||y_{ij} - \bm{U}_i||_2^2}{\sigma^2_i} - \log \left(2\pi\sigma_i^2\right) 
\end{align}
%\al{Chris, can we transform the density without considering the determinant of the Jacobian?}
%\chris{There's no need to worry about the Jacobian determinant here because we're not actually working in a transformed space for (8). It's simply a likelihood function relating the $U$'s to the model output $f$. Now, if the log likelihood was something more like $f(\cdot) - g(U)$, we'd need to take that into account.}
where $\bm{Y}$ represents the data of APs to fit and the $L^2$ norm as $||\cdot ||_2$. 

\subsection{Assumptions and limitations}

While this work is primarily a proof of concept to illustrate some of the challenges and advantages of CAP modeling within a spatial Bayesian context, there are major structural and distributional assumptions encoded within this model. The assumption of independent noise variates may be quite poor if there exists substantial unmodeled signal variance not captured by the MS model. In this case, the stochastic error is likely to be highly correlated, and the independent Gaussian noise model is a poor fit. A potential remedy in such a situation would be to include autocorrelated noise via an error model as commonly studied in time series analysis such as the AR$(p)$ family of probability models. It is also possible to imagine that the model errors are correlated across space and time; then, usage of a Gaussian process observational model would be appropriate. We note that this approach differs from the proposed GP prior usage in this work by employing a Gaussian process for part of the likelihood, rather than as a prior over model parameters.\\
%\al{What does it mean using a Gaussian process for part of the likelihood?}
%\chris{Instead of having an IID Gaussian likelihood as in equation (8), the likelihood is actually a multivariate normal density, with the covariance given by a kernel function like we used for the prior. Then we can't write (8) as a sum of IID terms.}
\par
Another major assumption in this work is that we only use a spatial prior for a single parameter out of the five associated with the MS model; this was primarily for computational tractability and also to highlight the differences in posterior inferences resulting from a spatially correlated versus uncorrelated prior distributional assumptions. A more comprehensive analysis would be to use a five-dimensional multivariate Gaussian process to model correlations between different parameters of the MS model as well as between instantiations of the same parameter at different locations in space.\\
\par
%@alejandro, please add more definitions to this table
%\al{Chris, I am not sure what $\mathcal{T}$ is. I do not define $\bm{x}_i$ as parameter of the model. I explain when and for how long we apply the stimulus in the Synthetic data generation section below. I also did not define $f_{\bm{\tau}}(\cdot)$ (at least not like that). It is defined at the beginning of the Hierarchical probability model section above. I think that we could combine this table with the priors section, or at least define it before because this notation is used before precisely when we introduce the priors for each parameter.}
\begin{table}[htbp]\caption{Notation}
\centering % to have the caption near the table
\begin{tabular}{c c p{10cm} }
\toprule
    $\bm{s}_1,...,\bm{s}_N$ & & Spatial coordinates of observational data\\
    $t_{ij}$ & &       Temporal coordinates for observational data \\
    $N$ & &                 Number of observation sites indexed by spatial coordinates \\
    $N_1$ & & Number of training locations\\
    $N_2$ & & Number of predicted locations \\
    $T$ & &                 Number of time points for each voltage signal   \\ 
    $U$ & & $N \times T$    matrix of voltage values observed over space and time  \\ 
    $\bm{\tau}$ & & Vector of site-specific parameters of MS\\
    $\bm{\lambda}^{(i)}$ & & Vector of non-site parameters of MS at position $i$\\
    $\mu_k$ & & Hierarchical mean of Mitchell-Schaeffer parameters pooling across locations \\
\bottomrule
\end{tabular}
\label{tab:notation}
\end{table}

% Given a system of differential equations $\bm{x}'(t)=\bm{f}(\bm{x},t,\bm{\theta})$, where the solution $\bm{x}=\bm{x}(t) \in {\rm I\!R}^n$ for each time $t \in [0,T],\, T \in {\rm I\!R}^+$, and $\bm{\theta} \in {\rm I\!R}^m$ is a vector of parameters, we use the following statistical model for the scalar variable representing the voltage:
% \[ y_{i}=x(t_i;\bm{\theta})+\epsilon_{i}.\]
% $y_{i}$ is the $i$-th observation of the individual being considered for the state $x(t_i;\bm{\theta}) \in {\rm I\!R}$ registered at time $t_{i}$, and $\epsilon_{i}$ is an independent normally distributed error with mean $\mu=0$ and standard deviation $\sigma$.\\
% \par

% Considering $\bm{\theta}$ as a random variable, by Bayes' theorem we have that 
% \[ p(\bm{\theta} | \bm{U}) \propto p(\bm{U} | \bm{\theta})p(\bm{\theta}),\]
% where $p(\bm{\theta}| \bm{U})$ is the final distribution (also known as the target or posterior distribution), $p(\bm{U} | \bm{\theta})$ is the likelihood, and $p(\bm{\theta})$ is the prior or initial distribution. 

\subsection{Parameter estimation}\label{parameterestimation}
%\al{Here $\bm{\theta}$ are all the parameters of hierarchical model, but theta was defined first as  $\bm{\theta}=(\tau_{out},\tau_{open},\tau_{close},u_{gate})$, the non-spatial parameters of MS. What is $\bm{X}$? }
%\chris{ "X" is generic notation for the data. If you want to use "Y" instead, that's fine.}
To estimate the parameters associated with the probability model of the previous section, we make use of Markov chain Monte Carlo methods for drawing samples from the posterior distribution with density $p(\bm{\theta}\vert \bm{Y}) \propto p(\bm{Y} \vert \bm{\theta}) \, p(\bm{\theta})$. As is standard in the calculation of Monte Carlo estimators for Bayesian inference, we construct a Markov chain $\bm{\theta}^{(1)},...,\bm{\theta}^{(j)}$ of $J$ samples drawn from the posterior distribution, with the posterior mean $\sum_{j=1}^L \bm{\theta}^{(j)}$ serving as our estimator of choice. In particular, we use a variant of Hamiltonian Monte Carlo, a Metropolis method that uses the gradient of the log posterior distribution distribution to construct a fictitious Hamiltonian system for which the canonical coordinates are given by the probability model parameters, and the energy is given by the negative log posterior.  \cite{neal2011}. This permits the use of the properties of Hamiltonian dynamics to efficiently sample from  $P(\bm{\theta|\bm{Y}})$. The system Hamiltonian function is defined in terms of parameter coordinates $\bm{\theta}$ and momenta variables $\bm{q}$ as 
\[H(\bm{\theta},\bm{q})=E(\bm{\theta})+K(\bm{q}),\] 
where $E(\bm{\theta})$ is a potential energy function linked to the posterior via generation of a Gibbs distribution as $P(\bm{\theta} \vert \bm{Y})=\frac{1}{Z} e^{-E(\bm{\theta})}$. Then,  $K(\bm{q})=\bm{q}^T\bm{\Sigma}^{-1}\bm{q}$ represents the kinetic energy; here $Z=\int_{\theta} e^{-E(\bm{\theta})} d\theta$ is a normalizing constant and $\bm{\Sigma}$ is a positive definite matrix called the mass matrix. This matrix rotates and rescales the sampler's velocity vector $\bm{q}$ to more naturally align with the geometry of the posterior distribution \cite{betancourt2018}. By simulating dynamics yielding samples of $\bm{\theta}$ and $\bm{q}$, we may then obtain marginal samples of $\bm{\theta}$ by discarding the values of $\bm{q}$.\\
\textbf{HMC}\\
    	Given $\bm{\theta} \in \mathbb{R}^M$ and $E(\bm{\theta})$,
	\begin{enumerate}
		\item Draw $\bm{q}$ from $P(\bm{q})=\exp(-K(\bm{q}))/Z_k$, $\bm{q}\sim \bm{\mathcal{N}}(0,\bm{I}_N)$
		\item $\bm{\theta}$ and $\bm{q}$ evolve according to 
		\begin{align*}
		\frac{d \theta_i}{dt} &= q_i, \\
		\frac{d q_i}{d t} &=-\frac{\partial}{\partial \theta_i} E(\bm{\theta}).
		\end{align*}
		\item Define $\epsilon$ and number of steps $l$ and solve
		\begin{align*}
		q_i\left(t+\frac{\epsilon}{2}\right)&=q_i(t)-\frac{\epsilon}{2} \frac{\partial}{\partial \theta_i} E(\bm{\theta}(t)),\\
	\theta_i(t+\epsilon)&=\theta_i(t)+\epsilon \, q_i\left(t+\frac{\epsilon}{2}\right), \\
	q_i(t+\epsilon)&=q_i\left(t+\frac{\epsilon}{2}\right)-\frac{\epsilon}{2} \frac{\partial}{\partial \theta_i} E(\bm{\theta}(t+\epsilon)),
	\end{align*}
	%This numerical method is conservative. 
to get $(\bm{\theta}^*,\bm{q}^*)$.
\item \[\bm{\theta}^{(t+1)}=\begin{cases}
    \bm{\theta}^{*} &  \text{ with probability } \alpha,\\
    \bm{\theta}^{(t)} &  \text{ with probability } 1-\alpha,
\end{cases}\]
where
\[\alpha=\min (1, \exp( H(\bm{\theta},\bm{q})-H(\bm{\theta}^*,\bm{q}^*)  ) ).\]
\end{enumerate}

The numerical integration scheme shown in step 3 above is called the leapfrog integrator, a finite-difference method to specifically designed to solve dynamical systems in classical mechanics with minimal error. To implement HMC, the parameters $\epsilon$ and $l$ need to be tuned, which in general is a challenging task. The No-U-Turn sampler (NUTS) avoids this need and is implemented in Stan, the statistical software that we used for our study. In NUTS, $l$ is determined adaptively at each iteration and there is also a procedure for adaptively setting $\Sigma^{-1}$ and step size $\epsilon$. The last step ($d$) accounts for the numerical error of the leapfrog algorithm where some proposals are accepted even when the error is not zero

\subsection{Synthetic data generation}

To assess the validity and usefulness of the previously described probabilistic model, we generated synthetic data using MS under the assumption that the parameter $\log \bm{\tau}$ is highly spatially correlated. In experiments, for an image obtained with optical mapping at a fixed time, each location is represented by a pixel. In real life, the locations are actually correlated since the electrical signals in the heart propagate by diffusion from one myocyte to the next. We fit MS to the last two APs in a series of 6 at $N_1$ locations which are represented on the grid by the vector $\bm{\tau}$. The rest of the parameters are not fixed but do not vary in space. 

We generated a spatial grid of $N=25, 100$ and $900$ locations with coordinates residing in the corresponding sets $\{1,...,5\}^2,\{1,...,10\}^2,\{1,...,30\}^2$. Each coordinate was assigned a raw independent standard normal noise variate; we then applied a Gaussian spatial filter to smooth the locations in space and thereby induce spatial correlations in the values of $\bm{\tau}$. We applied an affine transformation to the grid values in order to restrict them to be in the interval (0,1). A diagram for this setup is shown in Figures \ref{surface5}, \ref{surface10} and \ref{surface30}. We then selected  $N_1=4,6,14$ training locations respectively and drew values for the other model parameters applying Gaussian noise with $\sigma$ being 10\% of the values $\tau_{out},\tau_{open},\tau_{close},u_{gate}=(6, 150, 120, 0.13)$. These values, including the ones to generate the grid for $\tau_{in}$ are based on those found on the original MS paper \cite{mitchell2003} (where $\tau_{in}$=0.3). 

Because of the alternans behavior introduced in section \ref{CAP}, more than one CL are needed when fitting cardiac models to APs. Therefore, we simulated voltage time series for CLs paced at $350, 300$ and $276$ms at the $N_1$ training locations, and perturbed them with Gaussian noise with mean 0 and standard deviation equal to 0.03. Our objective with this synthetic data exercise was to determine whether or not recovery of the spatial parameter surface for $\bm{\tau}$ along with the system dynamics could be recovered given a limited number of voltage time series. The last two APs were taken in a series of 6---after reaching steady state---with only the last two CLs taken from the alternans regime and the first taken to the right of the BP. The resolution of the points selected from the APs to be fitted was of 0.5 ms for the first 4 ms and of 10 ms there after for each CL. 

%\al{Chris, I am not sure what this table is about; also, what should I add for T, all the times where the voltage is recorded?}
%\chris{Here's the layout for the table. The first column is going to give the name of the parameter like $\tau, \sigma^2$ or something similar. The second column will be the posterior mean calculated by taking the average over all of your draws. The third column will be the posterior standard deviation, calculated by taking the standard deviation over all of your draws. Remember to put the column names on the first row of the table.} \\

To summarize, we use a latent Gaussian process as a prior for the spatially varying parameter $\bm{\tau}$ and simulate data drawn from this prior. We then perform inference for a subset of $N_1=4,6$ or $14$ sampled locations---used as training points---and attempt to respectively predict the values of $\tau_{in}$ at the remaining $N_2=21,94$ or $886$ non-sampled locations employing Hamiltonian Monte Carlo to infer the model parameter distributions.

\subsection{Spatial parameter recovery}

To obtain the grid of predicted values for $\tau_{in}$, we took the mean of the $N_2$ posterior distributions for this parameter. Then we calculated the Pearson correlation coefficient between the true and predicted values at the $N_2$ locations to see if there was a linear relationship between them. The Pearson correlation coefficient is defined as 

\[ r_{xy}=\frac{\Sigma_{i=1}^n (x_i-\bar{x})(y_i-\bar{y})}{\sqrt{\Sigma_{i=1}^n (x_i-\bar{x})^2}\sqrt{\Sigma_{i=1}^n (y_i-\bar{y})^2}}.\]

To assess the validity of our method, for each CL used for fitting, we generated a population of action potentials at a desired location from the $N_2$ total locations, and compared them with the true APs at that specific location. To generate the population, we used the voltage model solution at the sample values from the posterior distribution of $\tau_{in}$ at the desired location and the four posterior distributions for the rest of the parameters at the higher level of the hierarchical model. To generate the true APs, we selected $\tau_{in}$ from the grid of true values at the desired location and the values we used to generate the synthetic data for the four other parameters. By doing so, there is no need to do inference at every location of the grid, which is computationally expensive. We take advantage of the correlated spatial structure on one of the parameters, derived from the heterogeneity of the heart observed through optical mapping. This task could have not be done by a non-spatial model since we would have only got the posterior distributions at the higher level, one for each one of the five MS parameters

\subsection{Implementation}
%\chris{Alejandro, here's the outline that you may want to follow for this section (1) say what statistical programming framework we used (Stan) and why we didn't write everything by hand. (2) Explain how Stan works at a high level (write down the forward model, let Stan figure out what the derivatives are), (3) explain how many samples and chains you ran, (4) explain what kind of computer you used, (5) provide a table of convergence diagnostic for each parameter ($\hat{R}$, effective sample size, and similar), and (6) describe some alternative software frameworks that you could use. PyMC is one example, another is Pyro or JAGS. This should take 2-3 paragraphs. Add on anything else which you think is relevant.}

%The Pearson correlation coefficient was 0.9759. 
%\chris{Alejandro, here you need to explain what the task is that we are proposing. Try writing a few paragraphs explaining (1) how the model might predict the spatial field at new locations, (2) why this might be a useful objective for assessment from a CAP modeling point of view, (3) why this task can't be done with a nonspatial model (hint - there's no spatial smoothing prior to tell us what the values should be at the held-out locations), (4) reasons for using the Pearson correlation coefficient between true and predicted (it's easy to understand and generically useful), (5) give the formula for the correlation coefficient. Don't give the correlation coefficient yet - save that for the results section.}

As mentioned in section \ref{parameterestimation}, HMC requires the two parameters $l$ and $\epsilon$, introduced by the use of the leapfrog integrator, to be tuned. Hoffman and Gelman \cite{Hoffman2014} created the No-U-Turn sampler (NUTS) to avoid the need of tuning the forementioned parameters and is implemented in Stan, the statistical platform that we used for our study. In NUTS, $l$ is determined adaptively at each iteration and there is also a procedure for adaptively setting both the mass matrix $\bm{\Sigma}^{-1}$ and the step size $\epsilon$. HMC also requires calculating the gradient of the target distribution, and this task is done in Stan through automatic differentiation. %\chris{It's not quite proper to call out institutions or people individually in the text within a scientific paper. I would recommend just deleting the clause '..it is maintained by a group of researchers at Columbia University} 
As for now, Stan has implemented several useful tools to assess different characteristics of the obtained samples, like the convergence of the chains through the $\hat{R}$ statistic, or the tail and bulk effective sample size (ESS), to mention a few. In Stan, it is also possible to run more than one chain depending on the number of cores of the computer used to run the programs. As alternatives, PyMC, Pyro or JAGS have NUTS implemented. 

Now, when working with differential equations, Stan has its own solvers, but we realized that by using our own solver, we saved computational time. We solved MS using an adaptive version of forward Euler with two step sizes for each CL. For the first 4 ms, corresponding to the upstroke (the stimulus was applied for the first ms), we used a step size of 0.1 ms, and a step size of 0.25 ms for the rest of the CL. We implemented our programs using one chain, the sample size for all cases was equal to 500 and the warm-up period was equal to 1000. We used RStan in R and ran the programs in an Apple MacBook Pro with M1 processor and 16 GB of unified RAM.

% We define a hierarchical model on the parameters of MS and simulate spatially correlated values on a two-dimensional domain for the parameter $\tau_{in}$, using a hierarchical but non-spatial prior structure for all other parameters (see Fig. \ref{hierarchical}). 

% To create a spatial grid for $\tau_{in}$ of size $n \times n$, we generated $n^2$ standard normally distributed numbers and we used a Gaussian filter to have them correlated. In reality, the values found in optical mapping recordings are a function of the action potential duration and those values are correlated since electricity in the heart propagates by diffusion from cell to cell (myocytes are connected through gap junctions). We then scaled and translate those numbers to have them centered around the value used for that parameter by Mitchell-Schaffer on their paper given a certain standard deviation. 

%Using a latent Gaussian process as a prior for the spatially varying parameter $\tau_{in}$, we perform inference in a subset of $N_1$ sampled locations used as training points to predict the values on the $N_2$ non-sampled locations, using NUTS in Stan for posterior inference..

\section{Results}
%%%% $5 \times 5$

\begin{figure}[]
	\centering
	\includegraphics[angle=0,width=1\textwidth]{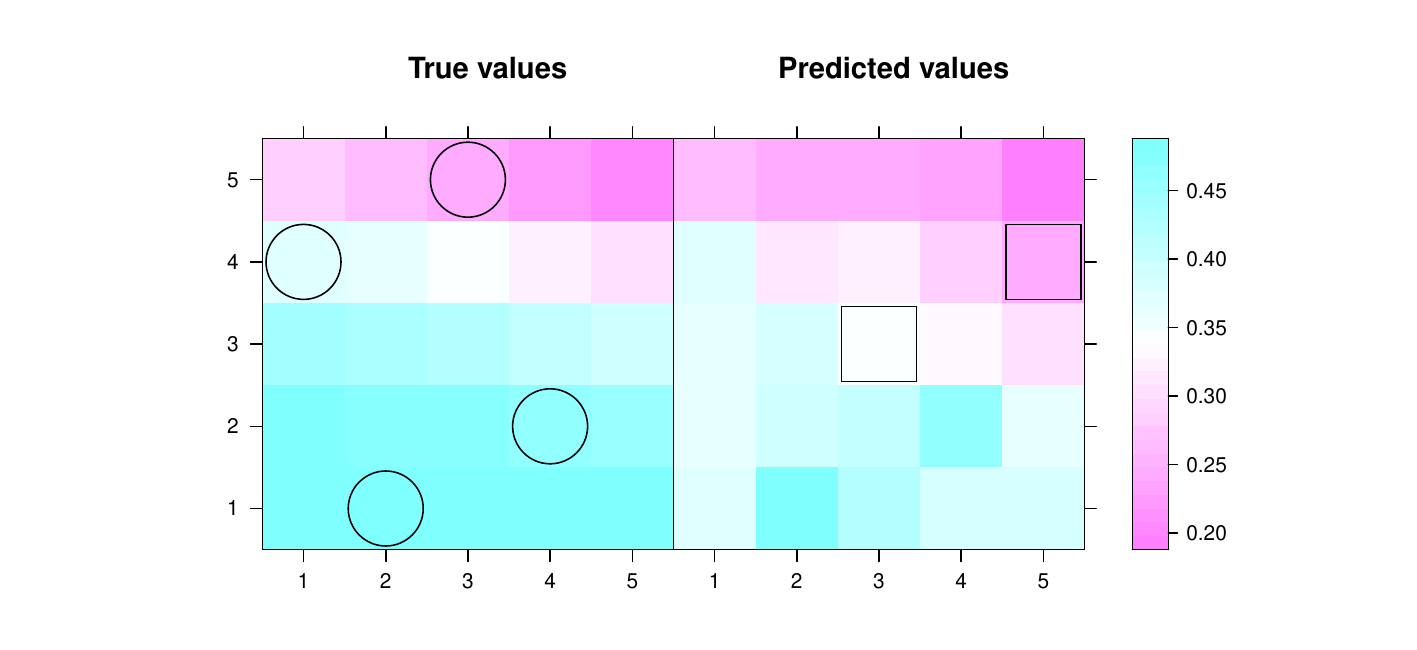}
	\caption{\label{surface5}Grid of true (left) and predicted values (right) for $\tau_1=\tau_{in}$ when the grid size was $5 \times 5$ and there were 4 training points (black circles). The squares point show the two training point for which we predicted the APs and BPs in Fig. \ref{5pos1and2}.}
\end{figure}

\begin{figure}[]
	\centering
	\includegraphics[angle=0,width=0.49\textwidth]{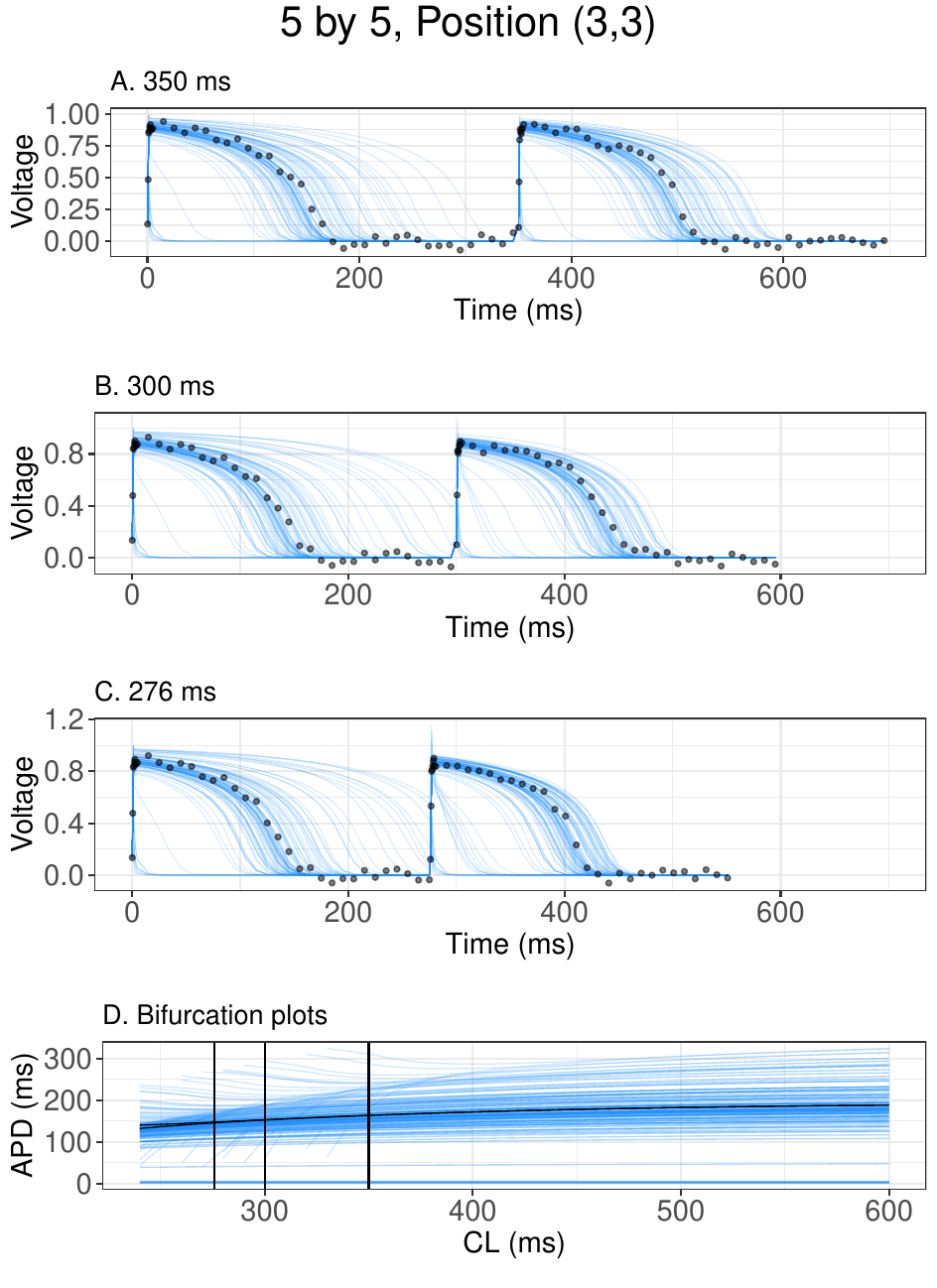}
	\includegraphics[angle=0,width=0.49\textwidth]{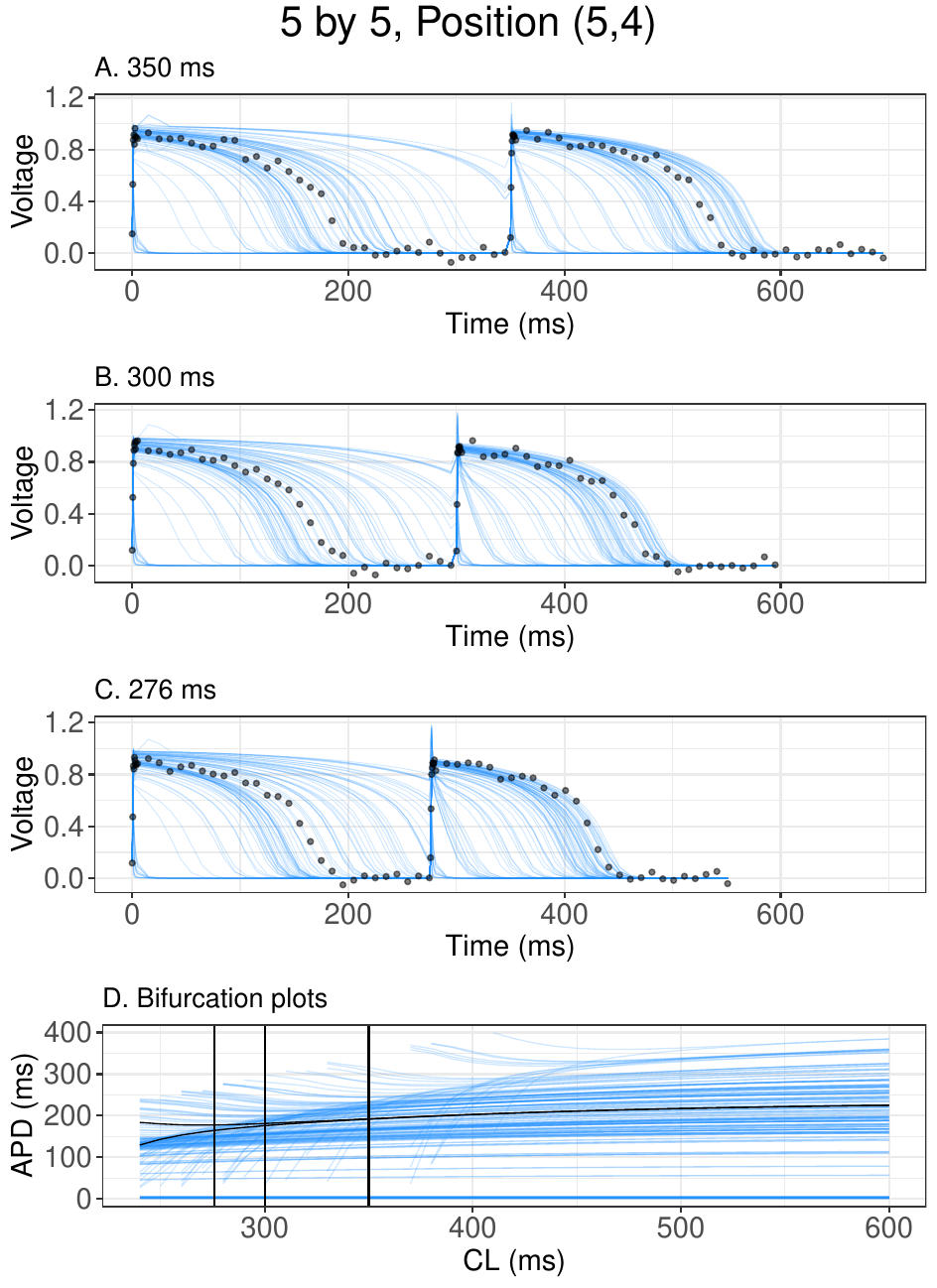}
	\caption{\label{5pos1and2} Comparison of true and predictive samples of action potential and bifurcation plots. Blue lines indicate samples from the noise-free posterior predictive distribution of the AP / BP values at new points in space. The black markers indicate the true values observed without noise. The size of the grid in this case is $5 \times 5$ and the predicted locations are $(3,3)$ and $(5,4)$.}
\end{figure}

In Fig. \ref{surface5} we observe the $5 \times 5$ grid of the true and predicted values the values of $\tau_{in}$ at each location. In this case, the four training points were selected at coordinates (2,1), (4,2), (1,4) and (3,5) for equal spacing and are represented by black circles. For illustrative purposes, we selected two of the test (i.e. non-training) $N_2$ positions, (3,3) and (5,4), as indicated by black squares in Fig. \ref{surface5} to generate 100 posterior predictive distributions of action potential and bifurcation plots for the cycle lengths of 350, 300, and 276 ms. used for fitting. We then compared these with the true APs represented by the black dots which were not used during inference. The results can be seen in Fig. \ref{5pos1and2}. In this figure, we observe that the posterior predictive distribution of APs exhibits substantial variation representing the posterior uncertainty in the parameters of the MS model at the new locations. Unsurprisingly, this dispersion is dampened for spatial locations which are closer to the coordinates used for training data. For the data corresponding to position (3,3), the true bifurcation plot does not show a pronounced bifurcation while some samples from the posterior predictive distribution do show a bifurcation. For position (5,4), the distribution displays more variation about true APs, and specifically for the 276ms CL, this predictive distribution of APs appears to systematically underestimate the true AP. 

In table \ref{table5} we present the mean posterior values, for each location in the case of $\tau_{in}$, for all the MS parameters next to the true values. We also show some metrics implemented in Stan to asses the convergence of the chains and to observe the bulk and tail effective sample size for this example.

%% 5 x 5

\begin{table}[]
\centering
\begin{tabular}{@{}ccccccccccccccc@{}}
\toprule
 Parameter  &  Position & True & Predictive mean& $\hat{R}$ & Bulk ESS & Tail ESS\\
\midrule
 $\tau_{in}$	&	(3,3)	&	0.421	&	0.338	&	1.001	&	476.726	&	505.271\\
	&	(5,4)	&	0.304	&	0.246	&	1.004	&	491.023	&	467.781\\
$\tau_{out}$	& N/A		&	6	&	6.082	&	0.998	&	478.265	&	483.113\\
$\tau_{open}$	& N/A		&	150	&	153.891	&	1.002	&	467.087	&	496.537\\
$\tau_{close}$	& N/A		&	120	&	115.319	&	1.005	&	450.483	&	459.604\\
$u_{gate}$	& N/A		&	0.13	&	0.124	&	1	&	518.237	&	411.188\\
\midrule
\end{tabular}
\caption{True and posterior mean estimates values and MCMC diagnostics  metrics for the $5 \times 5$ grid.}
\label{table5}
\end{table}

\begin{figure}[]
	\centering
	\includegraphics[angle=0,width=1\textwidth]{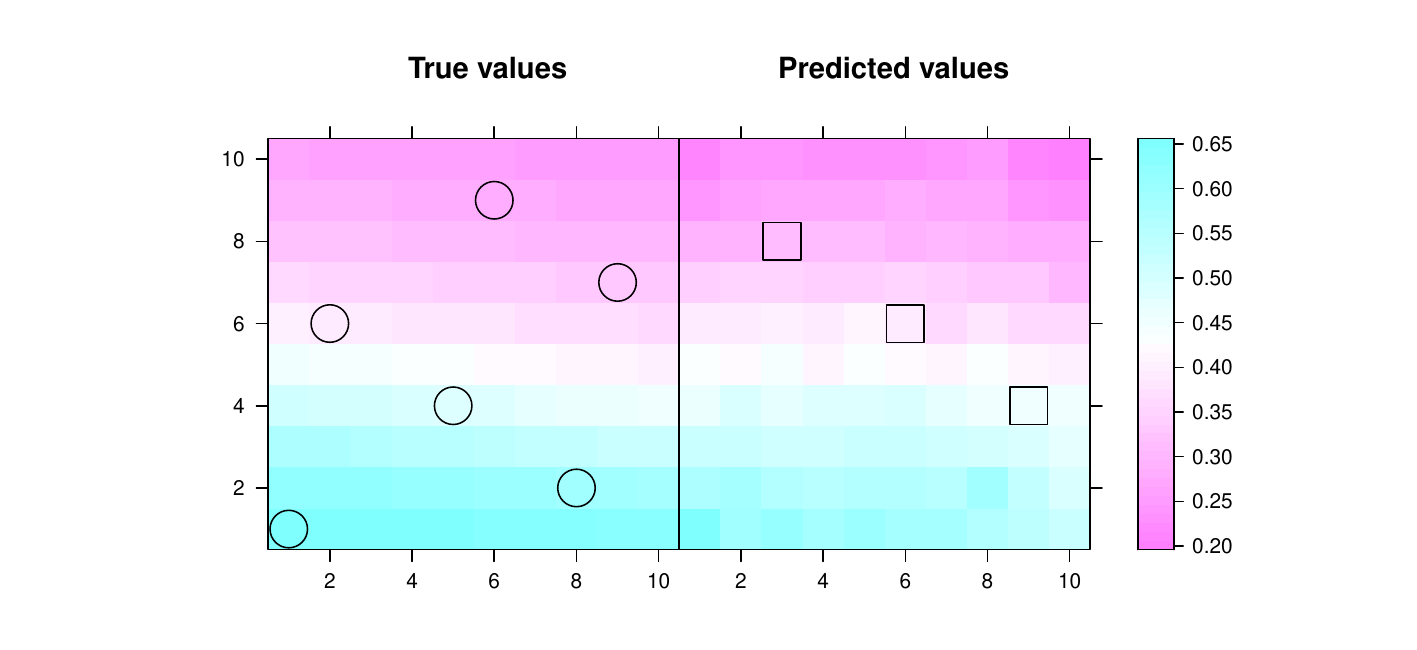}
	\caption{\label{surface10}Grid of true (left) and predicted values (right) for $\tau_1=\tau_{in}$ when the grid size was $10 \times 10$ and there were 5 training points (black circles). The squares point show the two training point for which we predicted the APs and BPs in Figures \ref{10pos1and2} and \ref{10pos3}.}
\end{figure}

\begin{figure}[]
	\centering
	\includegraphics[angle=0,width=0.49\textwidth]{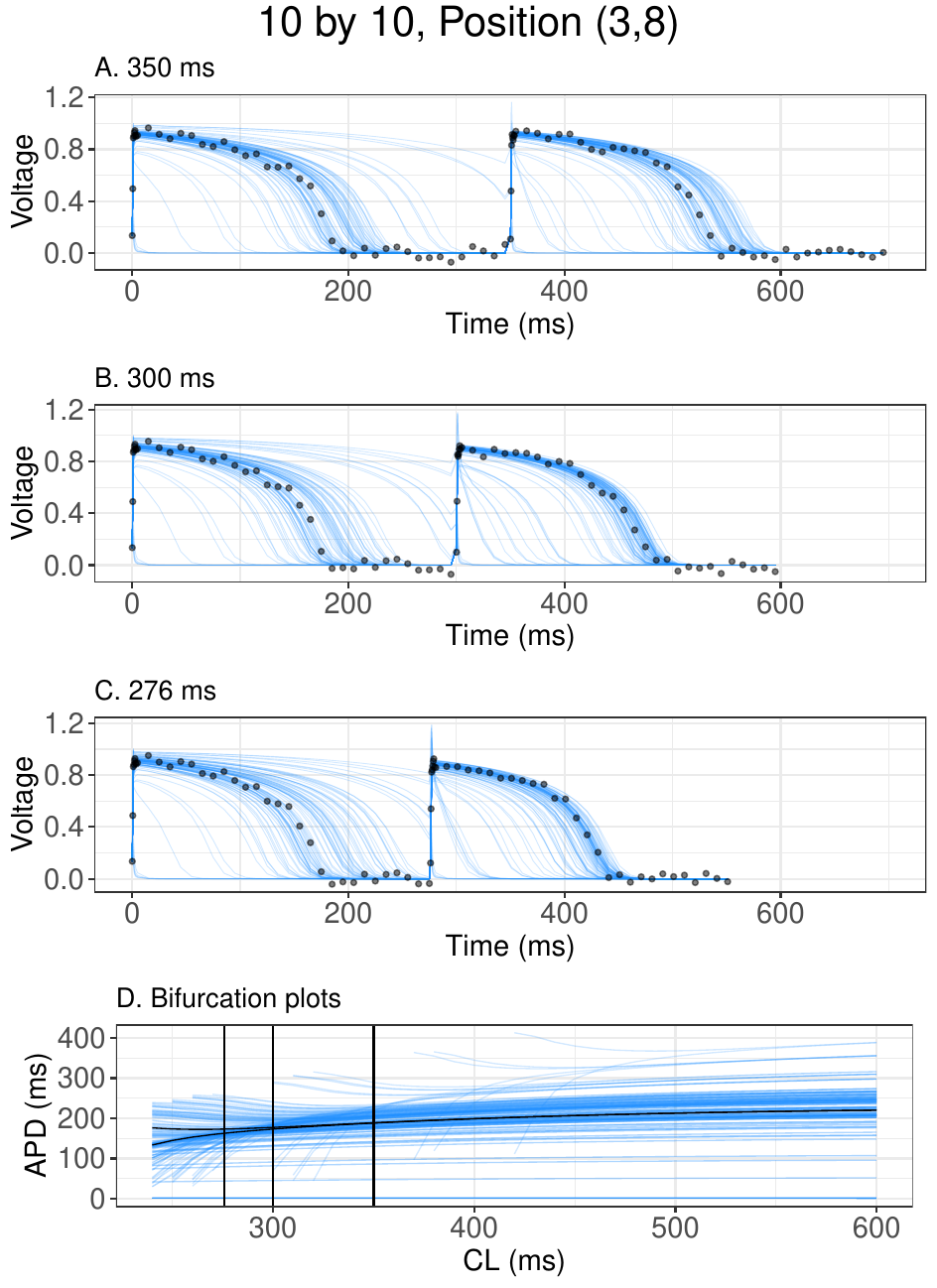}
	\includegraphics[angle=0,width=0.49\textwidth]{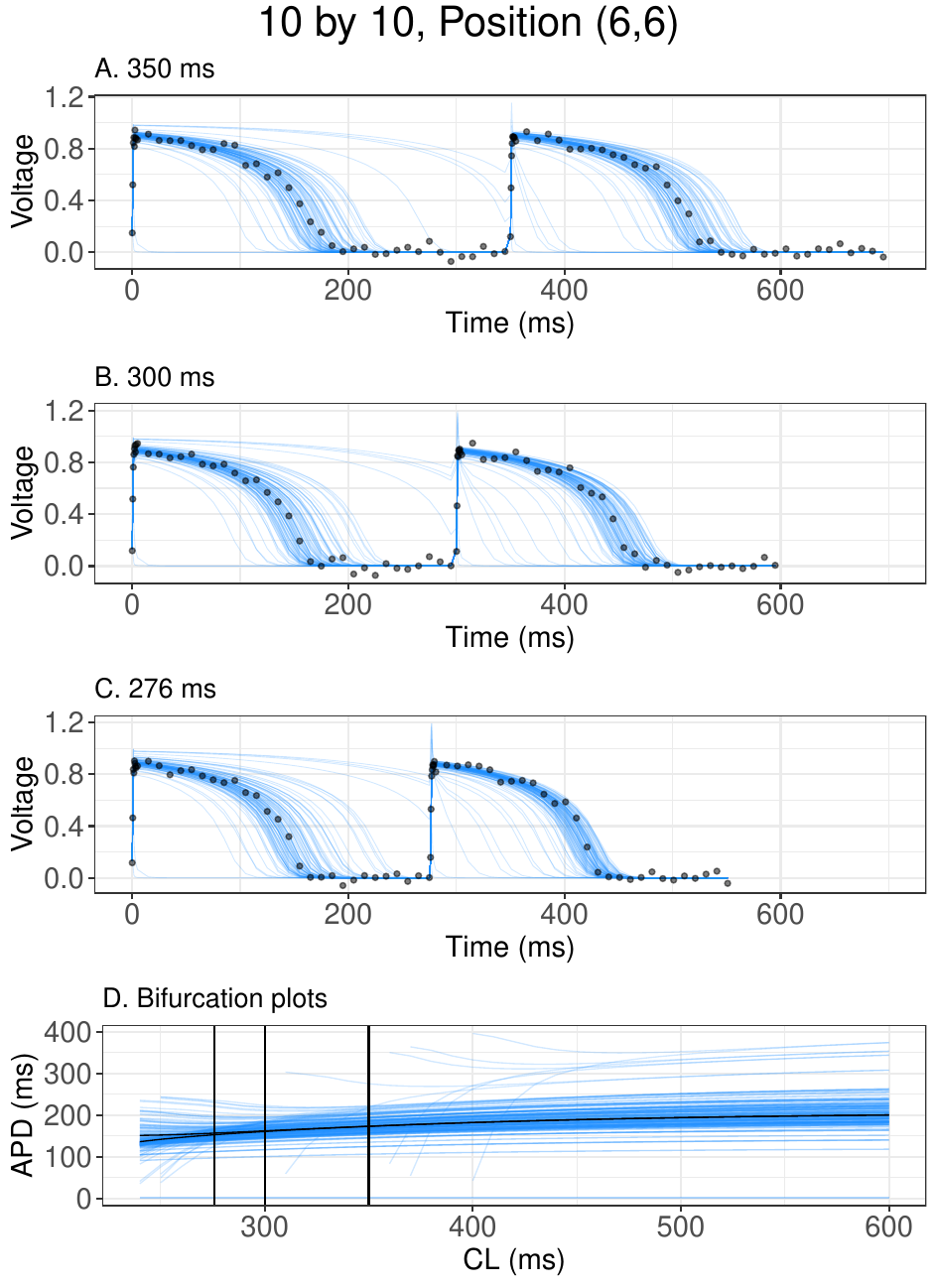}
	\caption{\label{10pos1and2} Predicted populations of APs and BPs (blue) along with the true AP to be predicted for the 3 CLs used for fitting. The size of the grid in this case us $10 \times 10$ and the predicted locations are (3,8) and (6,6).}
\end{figure}

\begin{figure}[]
	\centering
	\includegraphics[angle=0,width=0.5\textwidth]{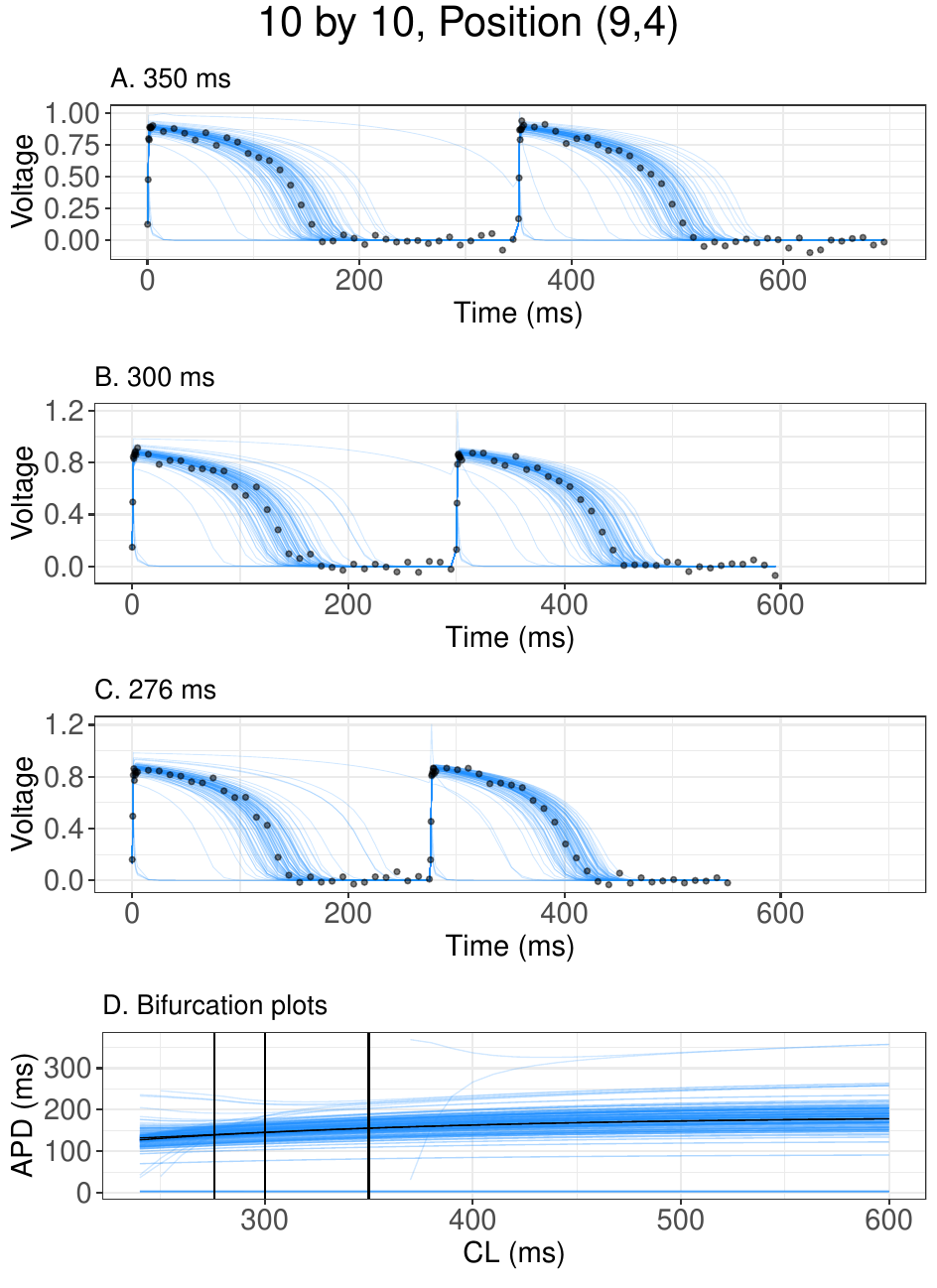}
	\caption{\label{10pos3} Predicted populations of APs and BPs (blue) along with the true AP to be predicted for the 3 CLs used for fitting. The size of the grid in this case us $10 \times 10$ and the predicted locations is (9,4).}
\end{figure}

In Fig. \ref{surface10} we see the $10 \times 10$ grid of the true and predicted values for each of the $\tau_{in}$ values at each location. The training points in this case ($N_1=6$) were (1,1), (8,2), (5,4), (2,6), (9,7) and (6,9), represented by circles. We selected 3 out the $N_2=94$ positions, (3,8), (6,6) and (9,4), represented by square, to generate the APs populations. As in the previous example, we also plotted a population of bifurcations plots to compare it with the true BP. The results are shown in Figures \ref{10pos1and2} and \ref{10pos3}.      

Once again, the populations are able to capture the shape and dynamics of the three selected positions. The populations of both, the APs and the BPs, present more variability for position (3,8), but the populations are again denser for the members that are closer to the true plots. Position (9,4) shows the least population variability of the three predicted positions. In Table \ref{table10} we show a table just like the one we showed in the previous example with the true and mean posterior predictive of the parameter values along with some Stan metrics.

%%%% $10 \times 10$

%% 10 x 10
\begin{table}[]
\centering
\begin{tabular}{@{}ccccccccccccccc@{}}
\toprule
Parameter  &  Position & True & Predictive mean& $\hat{R}$ & Bulk ESS & Tail ESS\\
\midrule
$\tau_{in}$	&	(3,8)	&	0.52	&	0.541	&	0.999	&	451.991	&	434.566	\\
$\cdot$	&	(6,6)&	0.493	&	0.507	&	0.998	&	400.749	&	514.387	\\
$\cdot$	&	(9,4)	&	0.463	&	0.466	&	0.998	&	400.749	&	514.387	\\
$\tau_{out}$	&	N/A	&	6.0	&	5.992	&	0.999	&	500.687	&	438.954	\\
$\tau_{open}$	&	N/A		&	150.0	&	147.418	&	0.998	&	586.925	&	483.624	\\
$\tau_{close}$	&	N/A		&	120.0	&	124.524	&	1.003	&	455.105	&	439.378	\\
$u_{gate}$	&		&	0.13	&	0.127	&	1.000	&	422.59	&	385.778	\\
\midrule
\end{tabular}
\caption{Real and mean posterior parameter values and some MCMC diagnostics for the $10\times10$ grid.}
\label{table10}
\end{table}

\begin{figure}[]
	\centering
	\includegraphics[angle=0,width=1\textwidth]{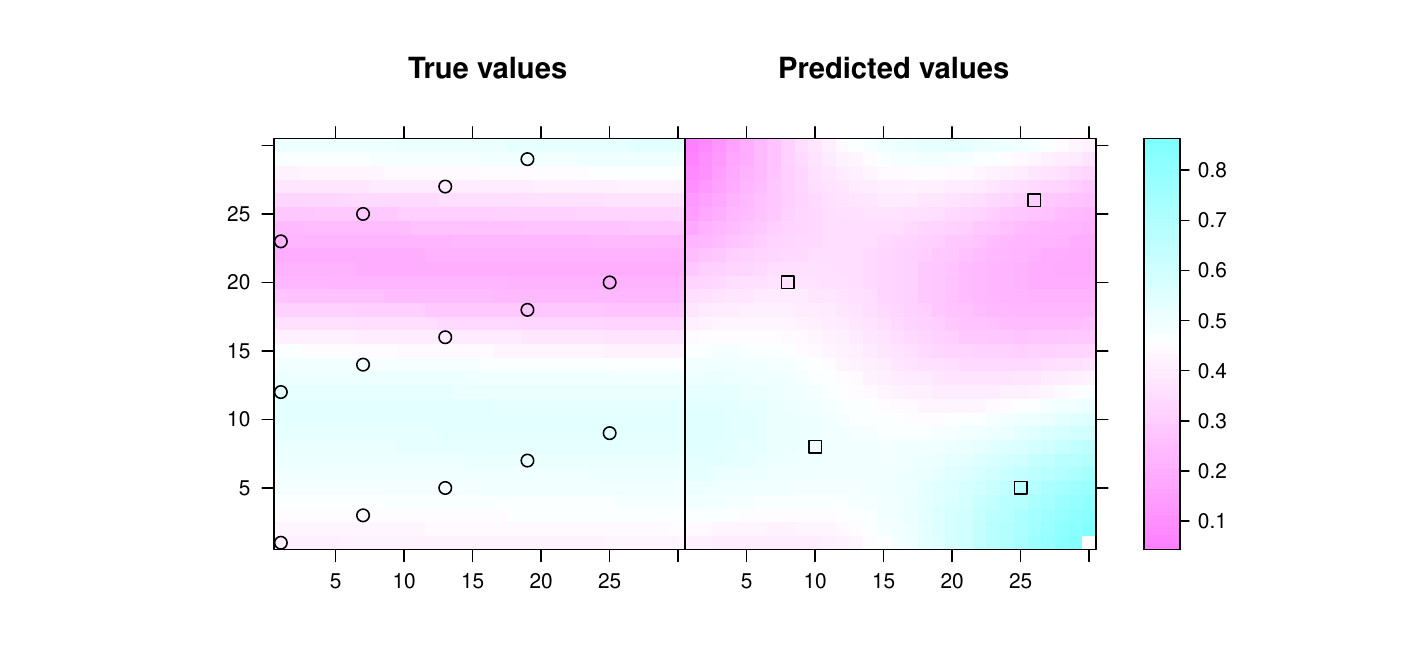}
	\caption{\label{surface30} Grid of true (left) and predicted values (right) for $\tau_1=\tau_{in}$ when the grid size was $30 \times 30$ and there were 14 training points (black circles). The squares point show the two training point for which we predicted the APs and BPs in Figures \ref{30pos1and2} and \ref{30pos3and4}.}
\end{figure}

\begin{figure}[]
	\centering
	\includegraphics[angle=0,width=0.49\textwidth]{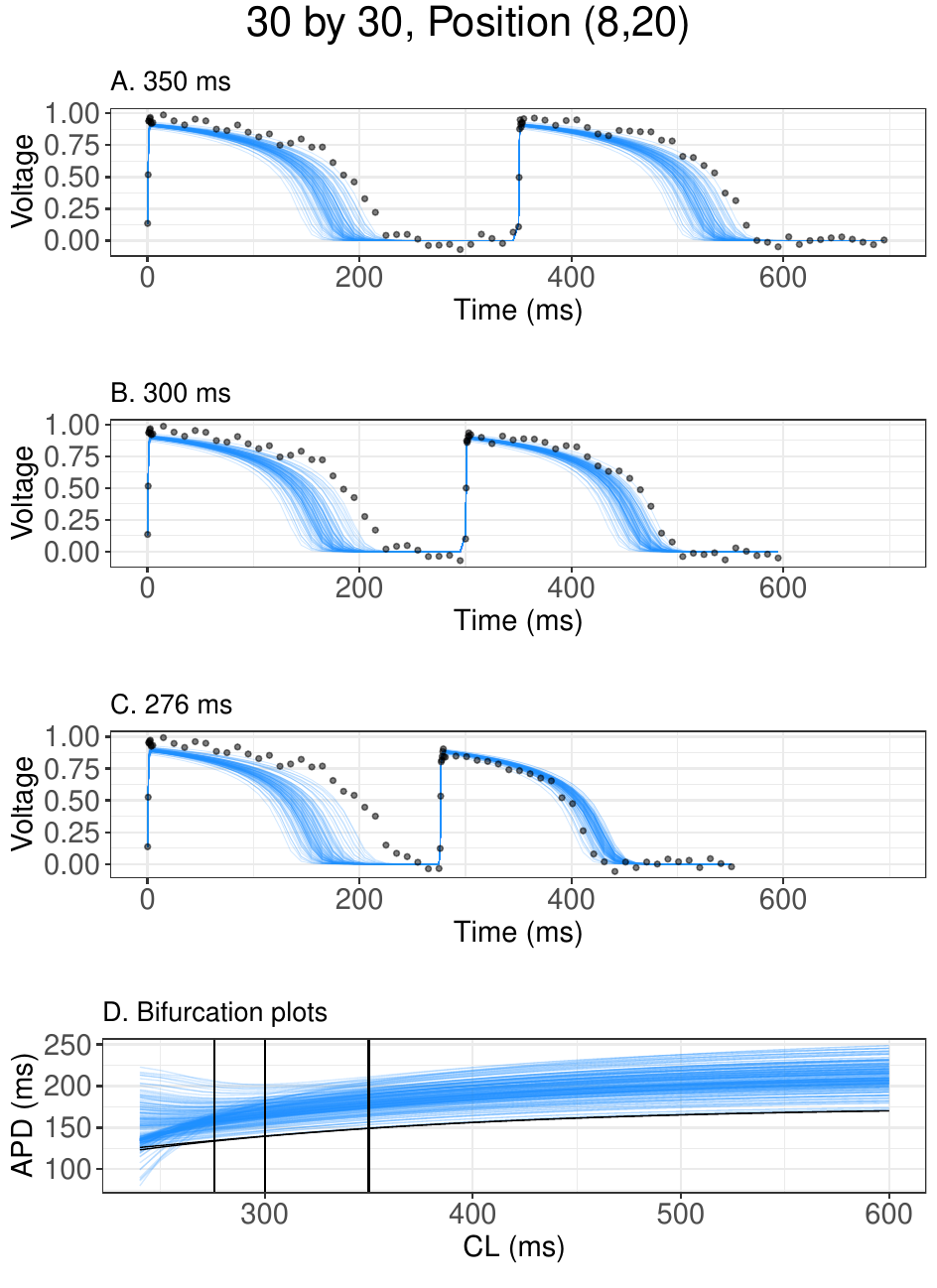}
	\includegraphics[angle=0,width=0.49\textwidth]{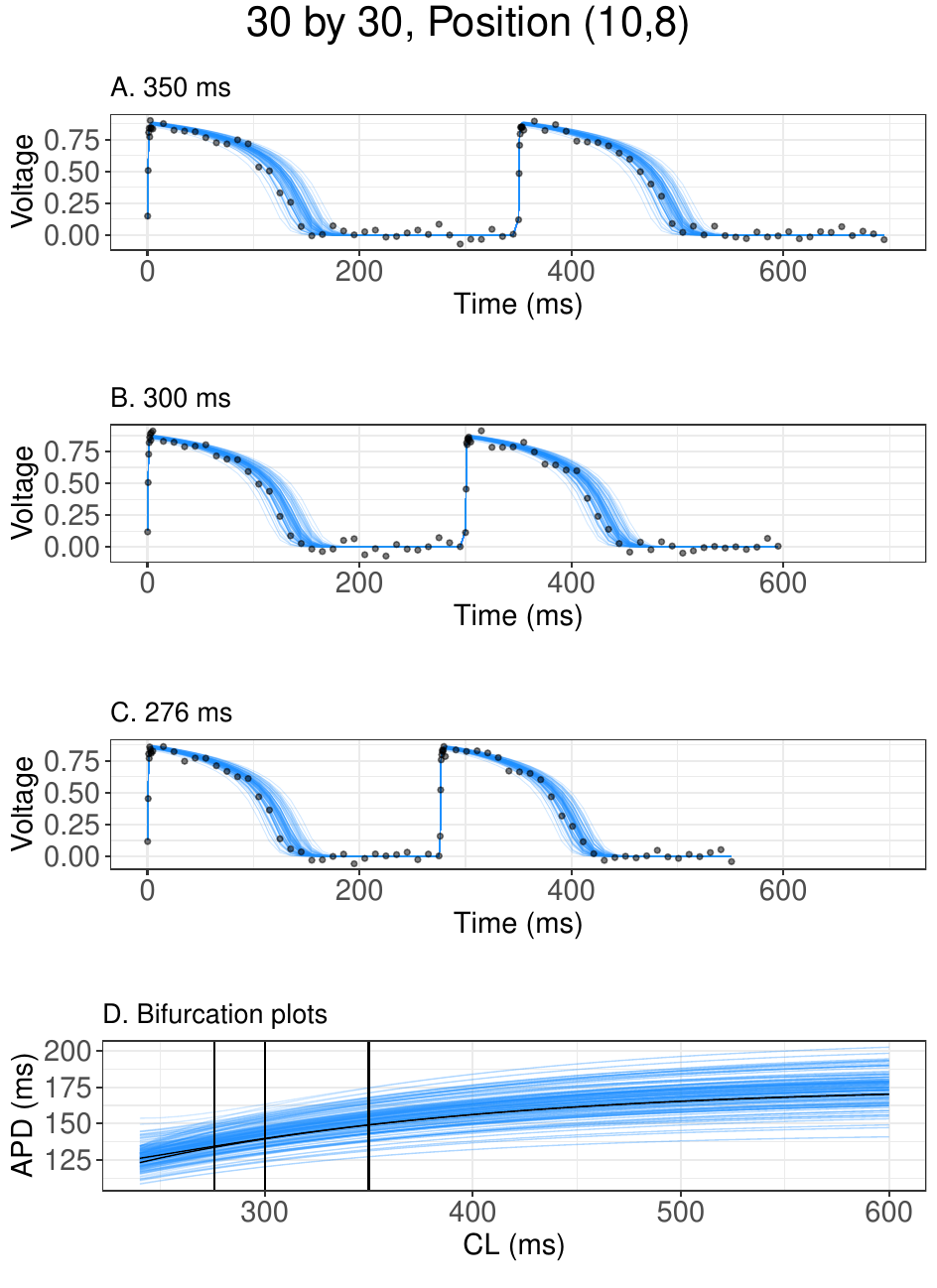}
	\caption{\label{30pos1and2} Predicted populations of APs and BPs (blue) along with the true AP to be predicted for the 3 CLs used for fitting. The size of the grid in this case us $30 \times 30$ and the predicted locations are (8,20) and (10,8).}
\end{figure}

\begin{figure}[]
	\centering
	\includegraphics[angle=0,width=0.49\textwidth]{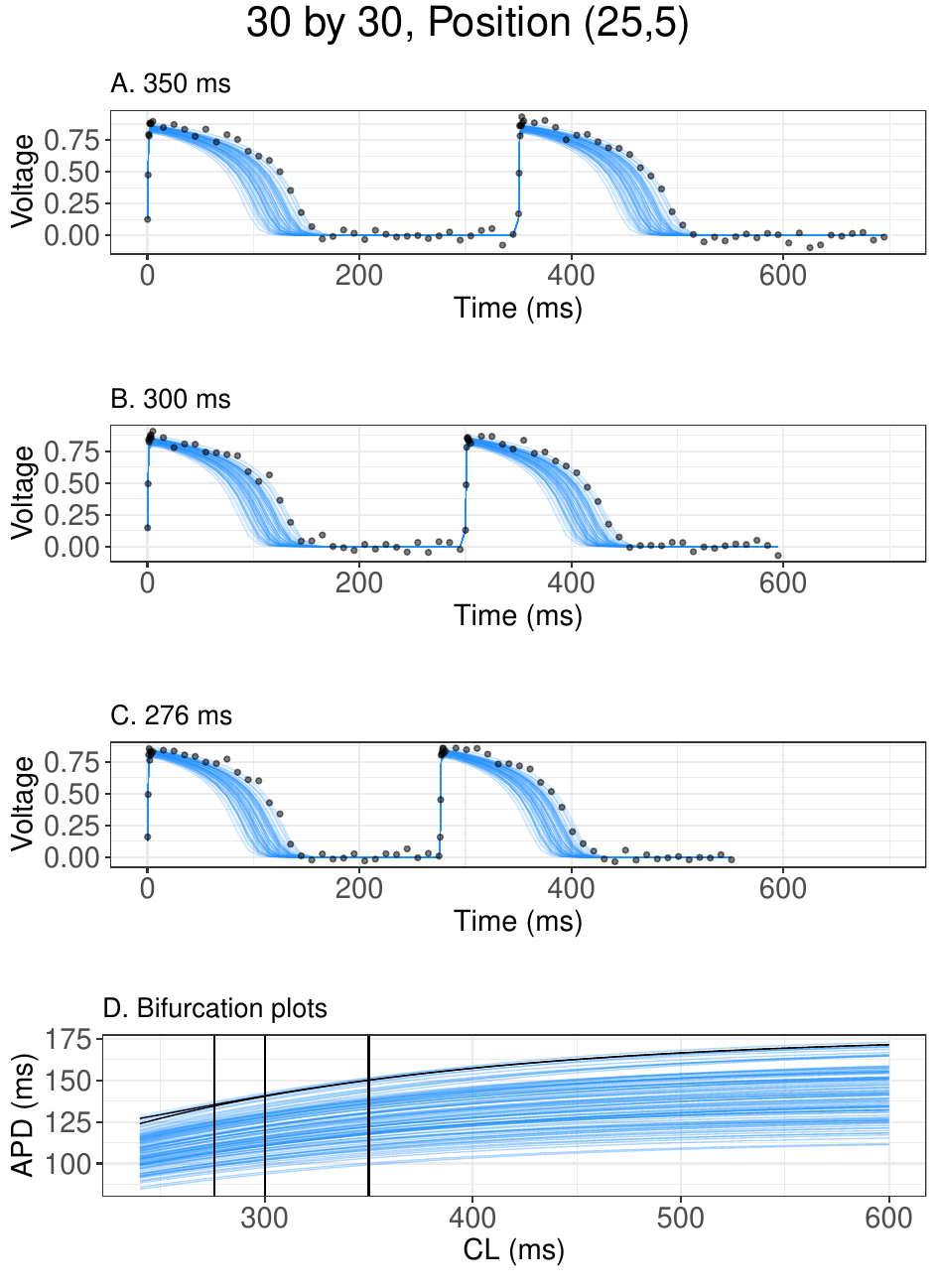}
	\includegraphics[angle=0,width=0.49\textwidth]{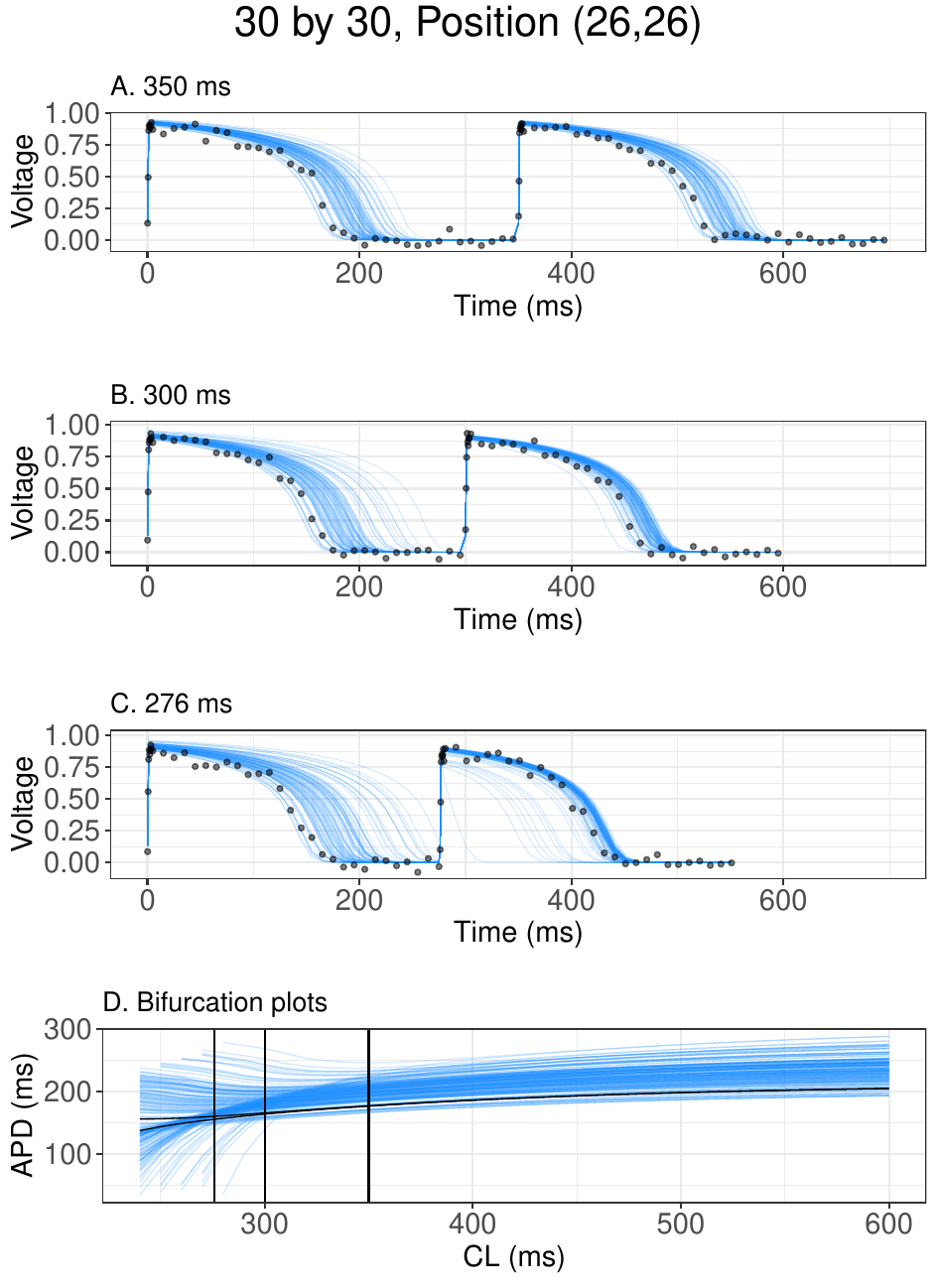}
	\caption{\label{30pos3and4} Predicted populations of APs and BPs (blue) along with the true AP to be predicted for the 3 CLs used for fitting. The size of the grid in this case us $30 \times 30$ and the predicted locations are (25,5) and (26,26).}
\end{figure}

Finally, in Fig. \ref{surface30} we show the $30 \times 30$ grid of the true and predicted values for each of the vales of $\tau_{in}$ at each location. The training points ($N_1=14$) were now (1,1), (7,3), (13,5), (19,7), (25,9), (1,12), (7,14), (13,16), (19,18), (25,20), (1,23), (7,25), (13,27) and (19,29), represented by circles. We selected 4 out the $N_2=986$ positions, (8,20), (10,8), (25,5) and (26,26), represented by squares, to generate the APs populations and compared them with the true APs for each CL. Again, we plotted a population of bifurcations plots to compare it with the true BP. The results are shown in Figures \ref{30pos1and2} and \ref{30pos3and4}.      

The results in this case are not as good as in the previous two examples, and we can corroborate that by just looking at the true and predicted grids of values. However, it is important to consider that in this case we only used 1.5 \% of all locations as training points (compared to the first and second examples, were the training points represented respectively, the 16 \% and 6\% of the total positions). The populations for position (8,20) are below the true APs and do not even reach the points that represent the fist action potential. The second action potential for the 276ms CL fits better the second AP compared to the other two CLs, even though for the first AP, this is also the worst fitting case. 

The true BP is also below all the members of the predicted population; as a matter of fact, the true BP does not seem to have a bifurcation, compared to the population of BPs that actually present a clear bifurcation point, the two branches starting around the 350ms value. Position (10,8) has the best fitting for both APs and BPs, and also the populations show the least variability. For position (25,5), the populations of APs and BPS are very close to the true plots but below the true one. In position (26,26) we see the opposite behavior: the populations of APs are close but above the true APs, except for some cases for the 276ms CL. The true BP seems to be more centered, but the bifurcation is least pronounced compared to the population of predicted BPs. 

In Table \ref{table30} we show again a table just like the one from the previous two examples, where we can see the mean posterior values, for each location in the case of $\tau_{in}$, for all the MS parameters next to the true values. Also the Stan metrics related to chain convergence and bulk and tail effective sample size for the $30 \times 30$ grid.

%%% 30 x 30
\begin{table}[]
\centering
\begin{tabular}{@{}ccccccccccccccc@{}}
\toprule
 Parameter  &  Position & True & Predictive mean& $\hat{R}$ & Bulk ESS & Tail ESS\\
\midrule
$\tau_{in}$	&	(8,20)	&	0.22	&	0.369	&	0.999	&	451.991	&	434.566	\\
	&	(10,8)	&	0.523	&	0.499	&	0.998	&	400.749	&	514.387	\\
	&	(25,5)	&	0.488	&	0.667	&	0.998	&	400.749	&	514.387	\\
	&	(26,26)	&	0.365	&	0.301	&	0.998	&	400.749	&	514.387	\\
$\tau_{out}$	&		&	6	&	5.992	&	0.999	&	500.687	&	438.954	\\
$\tau_{open}$	&		&	150	&	147.418	&	0.998	&	586.925	&	483.624	\\
$\tau_{close}$	&		&	120	&	124.524	&	1.003	&	455.105	&	439.378	\\
$u_{gate}$	&		&	0.13	&	0.127	&	1	&	422.59	&	385.778	\\
\midrule
\end{tabular}
\caption{Real and mean posterior parameter values and some Stan metrics for the $30 \times 30$ grid.}
\label{table30}
\end{table}

\begin{figure}[]
	\centering
	\includegraphics[angle=0,width=0.32\textwidth]{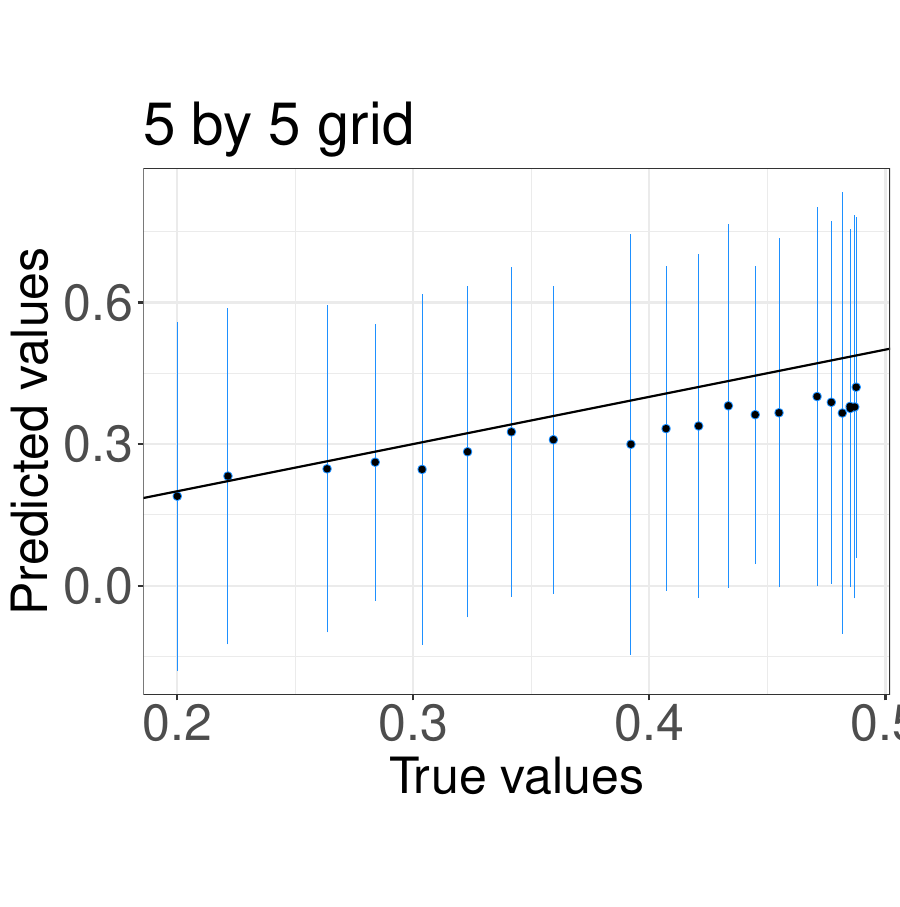}
	\includegraphics[angle=0,width=0.32\textwidth]{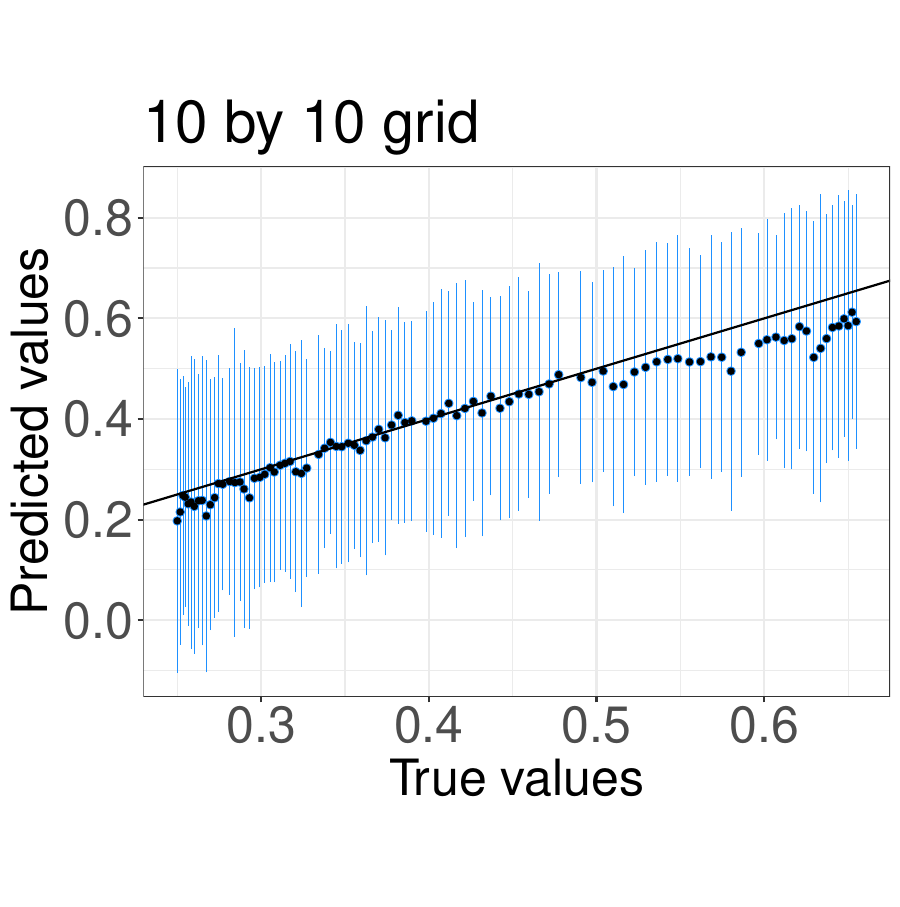}
	\includegraphics[angle=0,width=0.32\textwidth]{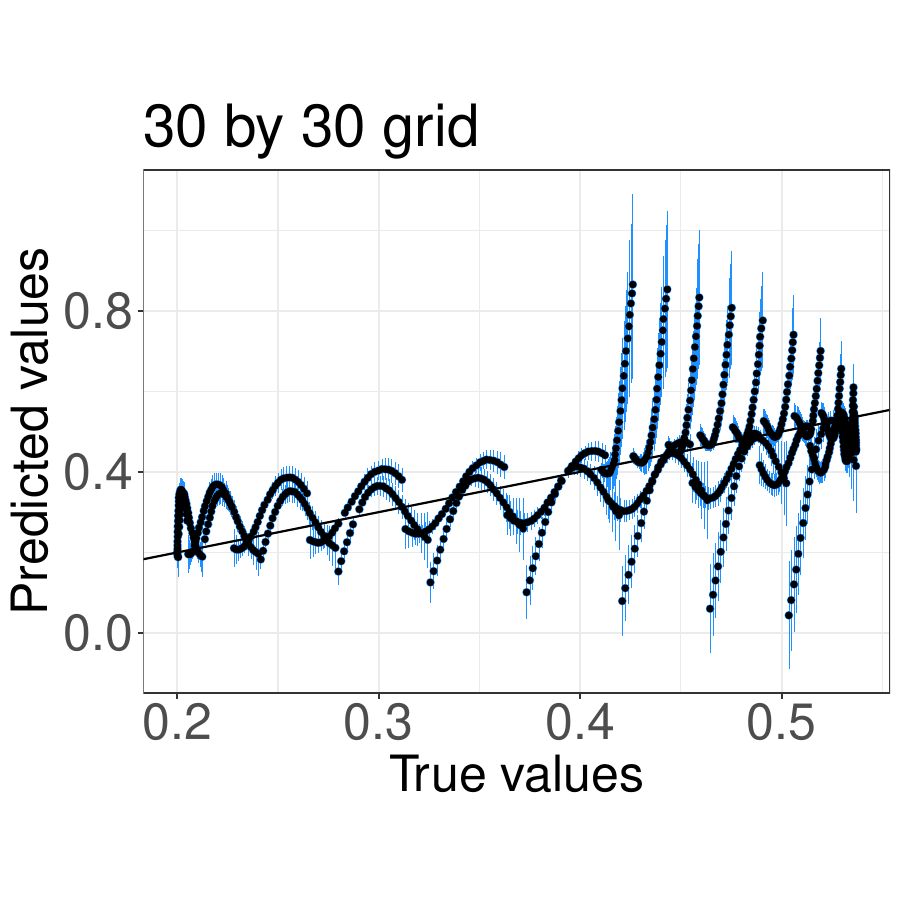}
	\caption{\label{correlation} Correlation between predicted vs. true values for $\tau_{in}$. The blue bars represent the standard deviation of the mean posterior predictive values and the back line represents the identity line.}
\end{figure}

The Pearson correlation coefficient ($R$) for each grid size can be seen in Table \ref{times} along with the computational time of the examples previously shown. We also plotted the predicted values of $\tau_{in}$ against the true values. We took the mean of the posterior distributions at each predicted location and also calculated their standard deviation (Fig. \ref{correlation}). The values for the $5 \times 5$ and $10 \times 10$ grid are very close to 1, which indicates a strong linear relationship between the true and the predicted values. This can also be seen when compared with the identity line, also plotted as a reference for each grid size. The case for the $30 \times 30$ case is not as good, and the graph looks a little bit strange because the grid in this case is not as the previous two; because the previous grids had monotone gradients for the values of $\tau_in$, which is not the case of the $30 \times 30$ grid. However, many values after 0.4 are very far from the identity line, which means that the predicted values were not accurate for those cases. The results improve if we add more training points. 

\begin{table}[]
\begin{center}
\begin{minipage}{174pt}
\begin{tabular}{@{}c|cccccccccccccc@{}}
\hline
N & 25 & 100 & 900\\
$N_1$ & 4 & 6 & 14\\
$r_{xy}$ & 0.96 & 0.98 & 0.63 \\ 
time(h)& 0.6  & 1.8 & 11.5 \\
divergences &396 & 0 & 0\\
\end{tabular}
\caption{Computational time and Pearson correlation coefficients for the three sizes.}
\label{times}
\end{minipage}
\end{center}
\end{table}

\section{Discussion}

We successfully did inference and made predictions on grids of different sizes selecting a small number of training points (no more than 16\% of the total positions). The correlation between the predicted and true values was closer to one when the grid size was respectively smaller, but should be pointed out that the percentage of training points also decreased with respect to the grid size. Even for the $30 \times 30$ case, the populations of action potentials were acceptable at the locations shown, and in some cases, the dynamics were described even if not prefectly. 

Now, the results improved when the non-training points were given as input to calculate the covariance matrix for the square exponential kernel when doing inference (results not shown). The reason for our choice of not including the non-training points when building the covariance matrix and showing those as examples in this work is two-folded. On the one hand, the number of training points we used was close to being the worst case scenario as the grid size was gradually being incremented; on the other hand, we wanted to reduce the computational time and still get acceptable results.  
For instance, the computation time for the grid of size 100 decreased from 5.7 to 1.8 h when not giving the non-training points as input for the covariance matrix, where the number of training points represented 6\% of the total number of positions on the grid. 

Related to the priors we used, we selected non only folded normals for the $\tau_{out},\tau_{open},\tau_{close}$ and $u_{gate}$, but also gamma distributions, getting comparable results (not shown). Since we used a hierarchical model, we selected folded normals so the distributions of the parameters at the higher level represented the mean (and variance) for all the positions, since the two parameters of a gamma distribution represent shape and scale (or rate) but not directly the mean and variance of the random variable in question. As for the distance parameter $\rho$ of the kernel, we used uniform and inverse gamma priors, selecting the parameters of the priors in such a way that the support included the minimum and maximum distance of any two points on the grid. The results shown in the paper are the ones for which we got better results. Finally, even though we present results of the Gaussian process prior where the mean was taken as constant, we also gave folded normals as priors for the mean, obtaining similar results as those shown in the examples. However, by the time increase in the latter case.

While we intend this study as a first proof-of-concept towards the learning of richer spatial patterns via Bayesian inference, it is worthwhile indicating potential future directions of progress. To make this problem richer in terms of forward modeling, we may consider a monodomain model of MS-like partial differential equation dynamics in cardiac tissue \cite{coveney2021} which would be substantially more computationally involved, but overall posing no theoretical issues with regard to statistical inference. Future efforts may make use of recently developed general-purpose computational frameworks \cite{hu2020, stanziola2021} which greatly expedite the generation of adjoint solvers, i.e, backpropagation of gradient information through the system dynamics. This type of work may also be implemented in secondary software packages making use of such frameworks but targeted towards cardiac electrophysiology researchers \cite{rognes2017}.
	
\section{Conclusion}

We created a mathematical model of cardiac APs which was able to capture the spatio-temporal voltage variability found in cardiac tissue. Exploiting the spatial correlation structure of myocytes observed through optical mapping voltage measurements, we created synthetic data employing the MS model, varying one of the model parameters simulated in a correlated-in-space grid of points and adding Gaussian noise to the rest of the parameter values used to create the APs. 
We developed a novel probabilistic model implemented in Stan using a hierarchical model on the parameters not considered correlated in space, and a latent Gaussian process prior for the parameter represented on the grid. Using a small number of sparsely-spaced training points, our technique allowed for rigorous uncertainty quantification and prediction at new points in space not used for training, permitting for reconstruction of surface electrophysiology dynamics.

We look forward to potential future implementations of the proposed approach on spatially explicit cardiac voltage data obtained from optical mapping procedures. In such a scenario, it may be advantageous to extend the spatial GP formulation to include space-time correlations; the intricacies of this approach are reviewed in \cite{guttorp2013}. Another potential avenue is to employ a more complex electrophysiology model such as the Fenton-Karma model \cite{fenton1998}.

%\section{Conclusions and Future Work}

%\begin{itemize}
%\item That the method worked well. 
%\item Apply the same methodology for a bigger grid, for a more complex model like Fenton-Karma and using optical mapping data. 
%\end{itemize}
\section*{Acknowledgments}  
% This section is not numbered.
This study was supported partially by NSF grant no. CNS-2028677.

\newpage
\printbibliography
%\begin{thebibliography}{99}{ %\small

%\bibitem{Murphy2018} Murphy SL, Xu J, Kochanek KD, Arias E. Mortality in the United States, 2017. NCHS data brief, no 328. Hyattsville, MD: National Center for Health Statistics; 2018.
%\bibitem{Fenton 1998} Fenton, F. \& A. Karma. (1998)``Vortex dynamics in three-dimensional continuous myocardium with fiber rotation: Filament instability and fibrillation''. Chaos 8, 20-47.
%\bibitem{Hoffman2014} Hoffman, M. D., \& Gelman, A. (2014). The No-U-Turn sampler: adaptively setting path lengths in Hamiltonian Monte Carlo. J. Mach. Learn. Res., 15(1), 1593-1623.
%\bibitem{Mitchell2003} Mitchell, C. C., \& Schaeffer, D. G. (2003). A two-current model for the dynamics of cardiac membrane. Bulletin of mathematical biology, 65(5), 767-793.
%\bibitem{Neal2011} Neal, R. M. (2011) MCMC Using Hamiltonian Dynamics. In Brooks, S., Gelman, A., Jones, G., \& Meng, X. L. (Eds.). (2011). Handbook of Markov Chain Monte Carlo. CRC press.
%\bibitem{Stan2021} Stan Development Team (2021). RStan: the R interface to Stan. R package version 2.21.2. http://mc-stan.org/}
%\end{thebibliography}

\end{document}